\documentclass[12pt]{article}

\usepackage{epsfig}

\newcommand{\sect}[1]{\setcounter{equation}{0}\section{#1}}

\def\be{\begin{equation}}
\def\ee{\end{equation}}
\def\ba{\begin{eqnarray}}
\def\ea{\end{eqnarray}}

\title{{\bf Trace anomaly driven inflation}}
\author{
{\bf S.W. Hawking}\thanks{email: S.W.Hawking@damtp.cam.ac.uk},
{\bf T. Hertog}\thanks{email: T.Hertog@damtp.cam.ac.uk}, \\
DAMTP, Centre for Mathematical Sciences, University of
Cambridge \\ Wilberforce Road, Cambridge CB3 0WA, United Kingdom
\\ and \\ 
{\bf H.S. Reall}\thanks{email: H.S.Reall@qmw.ac.uk} \\
Physics Department, Queen Mary and Westfield College, \\ Mile End
Road, London E1 4NS, United Kingdom 
\\ \\ Preprint DAMTP-2000-92, QMW-PH/00-10}

\date{25 October 2000}

\begin{document}

\maketitle

\begin{abstract}

This paper investigates Starobinsky's model of inflation driven by the
trace anomaly of conformally coupled matter fields. This model does
not suffer from the problem of contrived initial conditions that
occurs in most models of inflation driven by a scalar field. The universe
can be nucleated semi-classically by a cosmological instanton that is much
larger than the Planck scale provided there are sufficiently many matter
fields. There are two cosmological instantons: the four sphere and a
new ``double bubble'' solution. This paper considers a universe
nucleated by the four sphere. 
The AdS/CFT correspondence is used to calculate the correlation
function for scalar and tensor metric perturbations 
during the ensuing de Sitter phase. The analytic structure of the
scalar and tensor propagators is discussed in detail.
Observational constraints on the model are discussed. 
Quantum loops of matter fields are shown to
strongly suppress short scale metric perturbations, which implies that
short distance modifications of gravity would probably not be
observable in the cosmic microwave background. This is probably true
for any model of inflation provided there are sufficiently many matter
fields. This point is illustrated by a comparison of anomaly driven 
inflation in four dimensions and in a Randall-Sundrum brane-world model.

\end{abstract}

\pagebreak

\tableofcontents

\pagebreak

\sect{Introduction}

Inflation \cite{guth:81} in the very early universe seems the only natural
explanation of many observed features of our universe, 
in particular
the recent measurements of a Doppler peak in the cosmic microwave background
fluctuations \cite{debernardis:00}. However, while it provides an
appealing explanation for several cosmological problems,
it provokes the natural question of why the conditions were such as to
start inflation in the first place.

The new inflationary scenario \cite{linde:82, albrecht:82} 
was proposed primarily to overcome the
problem of obtaining a natural exit from the inflationary era. In this model, 
the value of the scalar is supposed to be initially confined
to zero by thermal effects. As the universe
expands and cools these effects disappear,
leaving the scalar field miraculously exposed on a 
mountain peak of the potential. If the low temperature potential is
sufficiently flat near $\phi =0$ then slow roll inflation will occur,
ending when the field reaches its true minimum $\phi_{c}$.
This scenario seems implausible because a high temperature would
confine only the average or expectation value of the scalar to zero.
Rather than be supercooled to a state with $\phi \sim 0$ locally, the
field fluctuates and rapidly forms domains with $\phi$ near $\pm\phi_{c}$ 
The dynamics of the phase transition is governed by
the growth and coalescence of these domains and not by a classical roll
down of the spatially averaged field $\phi$ \cite{mazenko:85}.
Because this and other problems, new inflation was largely abandoned
in favor of chaotic inflation \cite{linde:83} in which it is just
assumed that that the scalar field was initially displaced from 
the minimum of the potential.
One attempt to explain these initial conditions for inflation
in terms of quantum fluctuations of the
scalar field seems to lead to eternal inflation at the Planck 
scale \cite{linde:96}, 
at which the theory breaks down.
Another attempt, using the Hartle-Hawking ``no boundary'' proposal
\cite{hartle:83}, found that the most probable universes did not have
enough inflation \cite{hawking:98}. No satisfactory answer
to the question of why the scalar field was initially displaced from
the minimum of its potential has been found.

In this paper we will reconsider an earlier model, in which inflation
is driven by the trace anomaly of a large number of matter fields. The
Standard Model of particle physics contains nearly a hundred
fields. This is at least doubled if the Standard Model is
embedded in a supersymmetric theory. Therefore there were certainly a
large number of matter fields present in the early universe, so the
large $N$ approximation should hold in cosmology, even at the beginning of the
universe. In the large $N$ approximation, one performs the path
integral over the matter fields in a given background to obtain an
effective action that is a functional of the background metric:
\be
 \exp(-W[{\bf g}]) = \int d[\phi] \exp (-S[\phi; {\bf g}]).
\ee
One then argues that  the
effect of gravitational fluctuations is small in comparison to the
large number of matter fluctuations. Thus one can neglect graviton
loops, and look  for a stationary point of the combined gravitational
action and the effective action for the matter fields. This is
equivalent to solving the Einstein  equations with the source being
the expectation value of the matter energy  momentum tensor:
\be
 R_{ij} - \frac{1}{2} R g_{ij} = 8\pi G \langle T_{ij} \rangle,
\ee
where
\be
 \langle T^{ij} \rangle = -\frac{2}{\sqrt{-g}}\frac{\delta W}{\delta
g_{ij}}.
\ee
Finally, one can calculate linearized metric fluctuations about this 
stationary point metric and check they are small. This is confirmed 
observationally by measurements of the cosmic microwave background, 
which  indicate that the primordial metric fluctuations were of the 
order of $10^{-5}$ \cite{bennett:96}.

Matter fields might be expected to become effectively conformally
invariant if their masses are negligible compared to the spacetime
curvature. Classical conformal invariance is broken at the quantum
level \cite{capper:74} (see \cite{birrell:82, duff:94} for reviews),
leading to an anomalous trace for the energy-momentum tensor:
\be
 g^{ij} \langle T_{ij} \rangle \ne 0.
\ee
This trace is entirely geometrical in origin and therefore independent
of the quantum state. In a maximally symmetric spacetime, 
the symmetry of the vacuum implies
that the expectation value of the energy momentum tensor can be
expressed in terms of its trace
\be
 \langle 0| T_{ij} |0 \rangle = \frac{1}{4} g_{ij} g^{kl}
\langle 0| T_{kl} |0\rangle.
\ee
Thus the trace anomaly acts just like a cosmological constant for
these spacetimes. Hence a positive trace anomaly permits a de Sitter
solution to the Einstein equations \cite{dowker:76}. 

This is very interesting from the point of view of
cosmology, as pointed out by Starobinsky \cite{starobinsky:80}. 
Starobinsky showed that the de Sitter solution
is unstable, but could be long-lived, and decays into a matter
dominated Friedman-Robertson-Walker (FRW) universe. The purpose of
Starobinsky's work was to demonstrate that quantum effects of 
matter fields might resolve the Big Bang singularity\footnote{Another
paper \cite{fischetti:79}
which discussed the effects of the trace anomaly in cosmology
failed to obtain non-singular solutions because it included a
contribution from a classical fluid.}.
From a modern perspective, it is more
interesting that the conformal anomaly might have been the source of a
finite but significant period of inflation in the early universe. This
inflation would be followed by particle production and (p)reheating
during the subsequent matter dominated phase. Starobinsky's work is
reviewed and extended by Vilenkin in \cite{vilenkin:85}. For a more recent
discussion of the Starobinsky model, see \cite{fabris:00}.

Starobinsky showed that the de Sitter phase is unstable both to the
future and to the past, so it was not clear how the universe could
have entered the de Sitter phase. However, this problem can be
overcome by an appeal to quantum cosmology, which predicts that the
de Sitter phase of the universe is created by semi-classical
tunneling from nothing. This process is mediated by a four sphere 
cosmological instanton \cite{vilenkin:85}. One of the results of
this paper is that the four sphere is not the only cosmological
instanton in this model.

In order to test the Starobinsky model, it is necessary to compare its
predictions for the fluctuations in the cosmic microwave background
(CMB) with observation. This was partly addressed by Vilenkin
\cite{vilenkin:85}. Using an equation derived by Starobinsky
\cite{starobinsky:81}, Vilenkin showed that the amplitude of long
wavelength gravitational waves could be brought within observational
limits at the expense of some fine-tuning of the coefficients
parameterizing the trace anomaly. Density perturbations were discussed
by Starobinsky in \cite{starobinsky:83}. 

The analysis of Starobinsky and Vilenkin was complicated by the
fact that tensor perturbations destroy the conformal flatness of a FRW
background, making the effective action for matter fields hard to
calculate. However, we now have a way of calculating the effective
action for a particular theory, namely ${\cal N}=4$ $U(N)$ super
Yang-Mills theory, using the AdS/CFT correspondence
\cite{maldacena:98}. In this paper we will calculate the effective
action for this theory in a perturbed de Sitter background. This
enables us to calculate the correlation function for metric
perturbations around the de Sitter background. We can then compare our
results with observations. The fact that we are using the ${\cal N}=4$
Yang-Mills theory is probably not significant, and we expect our
results to be valid for any theory that is approximately massless 
during the de Sitter phase. One might think that our results could 
shed light on the effects of matter interactions during inflation 
since AdS/CFT involves a strongly interacting field theory. However,
as we shall explain, our results are actually independent of the
Yang-Mills coupling.

Our calculations will be performed in Euclidean signature (on the four
sphere), and then analytically continued to Lorentzian de Sitter
space. The condition that all perturbations are regular on the four
sphere defines the initial quantum state for Lorentzian perturbations.
The four sphere instanton is much larger than the Planck scale (since
we are dealing with a large $N$ theory), so there is a clear cut
separation into background metric and fluctuation. 

We shall include in our action higher derivative counterterms,
which arise naturally in the renormalization of the Yang-Mills
theory. There are three independent terms that are quadratic in the
curvature tensors: the Euler density, the square of the Ricci scalar
and the square of the Weyl tensor. The former just contributes a
multiple of the Euler number to the action. Metric perturbations do
not change the Euler number, so this term has no effect. The square of
the Ricci scalar has the important effect of adjusting the coefficient
of the $\nabla^2 R$ term in the trace anomaly. It is precisely this
term that is responsible for the Starobinsky instability, so by
varying the coefficient of the $R^2$ counter term we can adjust the
duration of inflation. The Weyl-squared counterterm does not affect
the trace anomaly but it can contribute to suppression of tensor
perturbations. The effects of this term were neglected by Starobinsky
and Vilenkin. They also neglected the effects of the non-local part of
the matter effective action. We shall take full account of all
these effects.
  
Vilenkin showed that the initial de Sitter phase is followed by a
phase of slow-roll inflation before inflation ends and the
matter-dominated phase begins. Since the horizon size grows
significantly during this slow-roll phase, it is important to
investigate whether modes we observe today left the horizon during the
de Sitter phase or during the slow-roll phase. If the present horizon
size left during the de Sitter phase, we find that the amplitude of metric
fluctuations can be brought within observational bounds if $N$, the
number of colours, is of order $10^5$. Such a large value for $N$ is
rather worrying, which leads us to the second possibility, that the
present horizon size left during the slow-roll phase. Our results then
suggest that the coefficient of the $R^2$ term must be at most of order
$10^8$, and maybe much lower, but $N$ is unconstrained (except by the
requirement that the large $N$ approximation is valid so that AdS/CFT 
can be used). We also find that the tensor perturbations can be suppressed
independently of the scalar perturbations by adjusting the coefficient
of the Weyl-squared counterterm in the action.

Inflation blows up small scale physics to macroscopic scales. This
suggests that inflation may lead to observational consequences of
small-scale modifications of Einstein gravity, such as extra
dimensions. However, we find that the non-local part of the matter 
effective action has the effect of strongly suppressing tensor
fluctuations on very small scales, a result first noted in flat space
by Tomboulis \cite{tomboulis:77}. This suggests that any small-scale 
modifications to four dimensional Einstein gravity would be
unobservable in the CMB since matter fields would dominate the graviton 
propagator at the scales at which such modifications might be expected 
to become important. This result is probably not restricted to trace
anomaly driven inflation since it is simply a consequence of the
presence of a large number of matter fields. As we have mentioned,
there really are a large number of matter fields in the universe and
these will suppress small-scale graviton fluctuations in any model of 
inflation. 

We illustrate this point by considering a Randall-Sundrum (RS)
\cite{randall:99} version of the Starobinsky model. In the RS model,
our universe is regarded as a thin domain wall in anti-de Sitter space
(AdS). RS showed that linearized four dimensional gravity is recovered on the
domain wall at distances much larger that the AdS radius of
curvature, but gravity looks five dimensional at smaller scales. Therefore,
if the AdS length scale is taken to be small, then the RS model is a
short distance modification of four dimensional Einstein gravity. We
shall show that when the large $N$ field theory is included, the
effects of the matter fields dominate the RS corrections to the
graviton propagator and render them unobservable. This work is an
extension of our previous paper \cite{hawking:00} to include the
effects of scalar perturbations and the higher derivative counterterms
in the action.

This paper is organized as follows. We start in section \ref{sec:O4} 
by showing that the
Starobinsky model has two instantons: the round four sphere and a new
``double bubble'' instanton. We consider only the four sphere
instanton in this paper. In section \ref{sec:pert} we use the AdS/CFT 
correspondence to calculate the effective action of the 
large $N$ Yang-Mills theory on a perturbed four sphere. Coupling this
to the gravitational action then allows us to compute the scalar and
tensor graviton propagators on the four sphere. In section
\ref{sec:analytic}, we discuss the analytic structure of our
propagators. The tensor propagator is shown to be free of
ghosts. In section \ref{sec:cont}, we show how our Euclidean
propagators are analytically continued to Lorentzian signature.
Section \ref{sec:observe} discusses two observational constraints
on the Starobinsky model, namely the duration of inflation and the
amplitude of perturbations. In section \ref{sec:RS}, we use the RS
version of the Starobinsky model as an example to illustrate how
matter fields strongly suppress metric perturbations on small scales.
Finally, we summarize our conclusions and suggest possible directions 
for future work.

\sect{$O(4)$ Instantons}

\label{sec:O4}

\subsection{Introduction}

Homogeneous isotropic FRW universes are obtained by analytic
continuation of cosmological instantons invariant under the action of
an $O(4)$ isometry group. In other words, we are interested in
instantons with metrics of the form
\be
\label{eqn:O4ans}
 ds^2 = d\sigma^2 + b(\sigma)^2 d\Omega_3^2.
\ee
We shall restrict ourselves to instantons with spherical topology, for
which $b(\sigma)$ vanishes at a ``North pole'' and a ``South
pole''. Regularity requires that $b'(\sigma)=\pm 1$ at these
poles. (Instantons with topology $R \times S^3$ also exist but are
excluded by the no boundary proposal because they are non-compact.)
The scale factor $b(\sigma)$ is determined by Einstein's equation\footnote
{We use a positive signature metric and a curvature convention for
which a sphere or de Sitter space has positive Ricci scalar.}
\be
 G_{ij} = 8\pi G \langle T_{ij} \rangle,
\ee
where the right hand side involves the expectation value of the energy
momentum tensor of the matter fields, which we are assuming to come from the
${\cal N}=4$ $U(N)$ super Yang-Mills theory. $\langle T_{ij} \rangle $
can be obtained for the
most general quantum state of the Yang-Mills theory consistent 
with $O(4)$ symmetry by using the trace anomaly and energy 
conservation, as we shall describe below.

\subsection{The trace anomaly}

The general expression for the trace anomaly of our strongly
coupled large $N$ CFT\footnote{
We shall often refer to the ${\cal N}=4$
Yang-Mills theory as a CFT even though it is not conformally invariant
on the four sphere.}
was calculated using AdS/CFT in \cite{henningson:98}. 
It turns out that it is exactly the same as the one loop result 
for the free theory, which is given for a general CFT 
by the following equation \cite{birrell:82,duff:94}
\be
\label{eqn:anomform}
 g^{ij} \langle T_{ij} \rangle = c F - a G + d \nabla^2 R
\ee
where $F$ is the square of the Weyl tensor:
\be
 F = C_{ijkl} C^{ijkl},
\ee
$G$ is proportional to the Euler density:
\be
 G = R_{ijkl} R^{ijkl} - 4 R_{ij} R^{ij} + R^2,
\ee
and the constants $a,c$ and $d$ are given in terms of the field
content of the CFT by
\be
 a = \frac{1}{360(4\pi)^2} \left( N_S + 11 N_F + 62 N_V \right),
\ee
\be
 c = \frac{1}{120(4\pi)^2} \left( N_S + 6N_F + 12 N_V \right), 
\ee
\be
 d = \frac{1}{180(4\pi)^2} \left (N_S + 6 N_F - 18 N_V \right),
\ee
where $N_S$ is the number of real scalar fields, $N_F$ the number of
Dirac fermions and $N_V$ the number of vector fields. The coefficients
$a$ and $c$ are independent of renormalization scheme but $d$ is
not. We have quoted the result given by zeta-function regularization
or point-splitting; the result given by dimensional regularization has
$+12$ instead of $-18$ as the coefficient of $N_V$ \cite{birrell:82}. 
In fact, $d$ can be adjusted to any desired value by adding the
finite counter term
\be
\label{eqn:counterterm}
 S_{ct} = \frac{\alpha N^2}{192 \pi^2} \int d^4 x \sqrt{g} R^2.
\ee
This counter term explicitly breaks conformal invariance. 
$\alpha$ is a dimensionless constant. The field content of
the Yang-Mills theory is $N_S = 6N^2$, $N_F = 2N^2$ (there are $4N^2$ Majorana
fermions, which is equivalent to $2N^2$ Dirac fermions) and $N_V =
N^2$. This gives
\be
 a = c = \frac{N^2}{64 \pi^2}, \qquad d = 0.
\ee
We have used the coefficient $-18$ for $N_V$ when calculating $d$ --
this is the value predicted by AdS/CFT \cite{henningson:98}. If $d=0$ then
inflation never ends in Starobinsky's model. We shall
therefore include the finite counter term, which does not change $a$ or $c$
but gives
\be
 d = \frac{\alpha N^2}{16 \pi^2}.
\ee
When we couple
the Yang-Mills theory to gravity, the presence of $S_{ct}$ implies that we are
effectively dealing with a higher derivative theory of gravity. It is,
of course, arbitrary whether one regards $S_{ct}$ as part of the
gravitational action or as part of the matter action. We have adopted the
latter perspective and therefore included an explicit factor of $N^2$
in the action (since there are ${\cal O}(N^2)$ fields in the
Yang-Mills theory).

\subsection{Energy conservation}

Having obtained the trace of the energy-momentum tensor, we can use
energy-momentum conservation to obtain the full energy-momentum
tensor. Introduce the energy density $\rho$ and pressure $p$, defined 
in an orthonormal frame by
\be
 \langle T_{\sigma \sigma} \rangle = -\rho, 
 \qquad \langle T_{\alpha\beta} \rangle
 = p \delta_{\alpha\beta}.
\ee
The minus sign in the first expression arises because we are
considering Euclidean signature. These must obey
\be
\label{eqn:rhop}
 -\rho + 3p = \langle T \rangle,
\ee
and we also have the energy-momentum conservation equation
\be
 \rho' + \frac{3}{b'}{b} (p+\rho) = 0.
\ee
Eliminating $p$ gives an equation for $\rho$:
\be
 \left(b^4 \rho\right)' = -b^3 b' \langle T \rangle.
\ee
Substituting in the expression for $\langle T \rangle$ and integrating
gives
\be
 \rho = \frac{3N^2}{8\pi ^2 b^4} \left[\frac{(1-{b'}^2)^2}{4} +
\alpha \left(b^2 b' b''' - \frac{1}{2} b^2 {b''}^2 + b {b'}^2 b'' -
\frac{3}{2} {b'}^4 + {b'}^2 \right) + C \right].
\ee
The expression for $p$ is easily determined from equation \ref{eqn:rhop}. 
The appearance of the constant of integration $C$ shows that the
quantum state can contain an arbitrary amount of radiation. Setting
$C=\alpha/2$ reproduces the energy-momentum tensor for the vacuum state. The
cosmology resulting from the trace anomaly in the presence of an arbitrary
amount of null radiation was investigated in \cite{fischetti:79}. The
cosmological solutions obtained were generically singular. However,
Starobinsky \cite{starobinsky:80} showed if this null radiation 
is not present (i.e., if $C=\alpha/2$) then non-singular solutions 
can be obtained. 

To conclude, we have found the energy-momentum tensor for a strongly
coupled large $N$ Yang-Mills theory in the most general quantum state that is
consistent with $O(4)$ symmetry. The effects of strong coupling do not
show up in our energy-momentum tensor, which is of the same form as
used in \cite{fischetti:79, starobinsky:80}. In the next subsection 
we shall use this result in the Einstein equations to determine 
the shape of the instanton.

\subsection{Shape of the instanton}

Taking the $\sigma\sigma$ component of the Einstein equation gives
\be
 G_{\sigma\sigma} \equiv 3 \frac{{b'}^2-1}{b^2} = -8 \pi G \rho.
\ee
Substituting in our result for $\rho$ gives
\be
\label{eqn:ein}
 \frac{1-{b'}^2}{b^2} = \frac{N^2 G}{\pi} \left[\frac{(1-{b'}^2)^2}{4
 b^4} + \alpha \left( \frac{b'b'''}{b^2} - \frac{{b''}^2}{2b^2} +
 \frac{{b'}^2 b''}{b^3} - \frac{3 {b'}^4}{2b^4} + \frac{{b'}^2}{b^4}
 \right) + \frac{C}{b^4} \right].
\ee
Regularity at the poles of the instanton requires $b' \rightarrow \pm
1$ as $b \rightarrow 0$. Substituting this into equation
\ref{eqn:ein}, one finds that $b''=0$ and $C=\alpha/2$ 
are also required for regularity at the
poles. In other words, the no boundary proposal has singled out a
particular class of quantum states for us, namely those that do not
contain any radiation. These are precisely the states that can give
rise to non-singular cosmological solutions. In our picture this is
because such cosmological solutions can be obtained from a Euclidean
instanton. 

It is convenient to introduce a length scale $R$ defined by 
\be
\label{eqn:Rdef}
 R^2 = \frac{N^2 G}{4 \pi}.
\ee
We can now define dimensionless variables
\be
 \tilde{\sigma} = \sigma/R, \qquad f(\tilde{\sigma}) = b(\sigma)/R.
\ee
Equation \ref{eqn:ein} becomes
\be
\label{eqn:fein}
 \frac{1-{f'}^2}{f^2} = \frac{(1-{f'}^2)^2}{f^4} + 2\alpha
\left(\frac{2 f' f'''}{f^2} - \frac{{f''}^2}{f^2} + 2\frac{{f'}^2
f''}{f^3} - 3\left(\frac{f'}{f}\right)^4 + 2\frac{{f'}^2}{f^4} +
\frac{1}{f^4} \right).
\ee 
The boundary conditions at the poles are $f=0$, $f'=\pm 1$, $f''=0$
(where a prime now denotes a derivate with respect to $\tilde{\sigma}$).
One solution to equation \ref{eqn:fein} is
\be
 f(\tilde{\sigma}) = \sin \tilde{\sigma},
\ee
which simply gives us a round four sphere instanton. Note that the
expression multiplying $\alpha$ vanishes for this solution. Another
simple solution is 
\be
 f(\tilde{\sigma}) = \tilde{\sigma},
\ee
i.e. flat Euclidean space.

In order to integrate \ref{eqn:fein} numerically, we assume that
$\tilde{\sigma}=0$ is a regular ``North pole'' of the instanton. 
We start the integration at $\tilde{\sigma} = \epsilon$. The boundary
conditions for the integration are
\be
 f(\epsilon) = \epsilon + \frac{1}{6} f'''(0) \epsilon^3 + \ldots
\ee
\be
 f'(\epsilon) = 1 + \frac{1}{2} f'''(0) \epsilon^2 + \ldots
\ee
\be
 f''(\epsilon) = f'''(0) \epsilon + \ldots
\ee
We shall neglect the higher order terms (denotes by the ellipses) in
our numerical integration. It is important to retain all of the terms
displayed in order to obtain $f'''(\epsilon) = f'''(0)+\ldots$ 
from the equation of motion. Note that $f'''(0)$ is a free parameter. 
Our strategy is to choose the value of $f'''(0)$ so that the instanton
is compact and closes off smoothly at the South pole. 

The instanton is non-compact when $f'''(0)>0$. The solution is flat
Euclidean space when $f'''(0)=0$. We shall therefore concentrate on
$f'''(0)<0$. The four sphere solution has $f'''(0)=-1$. It is
convenient to discuss the cases $\alpha >0$ and $\alpha <0$
separately.
\begin{figure}
\begin{picture}(0,0)
\put(28,0){\tiny{$0$}}
\put(88,0){\tiny{$1$}}
\put(146,0){\tiny{$2$}}
\put(203,0){\tiny{$3$}}
\put(261,0){\tiny{$4$}}
\put(18,55){\tiny{$0.2$}}
\put(18,103){\tiny{$0.4$}}
\put(18,150){\tiny{$0.6$}}
\end{picture}
\centering{\psfig{file=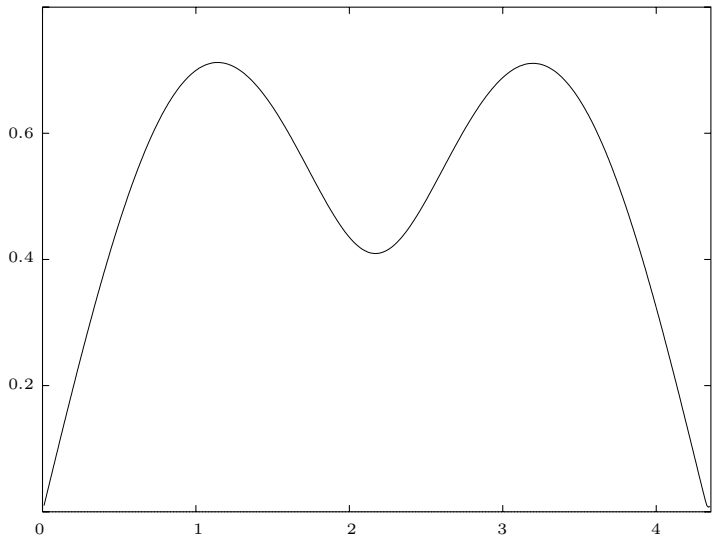,width=4.in}}
\caption{Scale factor $f(\tilde{\sigma})$ 
for a regular ``double bubble'' instanton with 
$\alpha=-1$ and $f'''(0)=-2.05$.}
\label{fig:reg}
\end{figure}

If $\alpha > 0$ then there are two types of behaviour.
(i) $-1 < f'''(0) < 0$ the instanton is
non-compact. For $f'''(0)$ close to $-1$, the scale factor increases
to a local maximum and then starts to decrease. However, before
reaching $f=0$, the scale factor turns around again and increases
indefinitely. (ii) $f'''(0)<-1$. These instantons are compact
but do not have a regular South pole since $b'$ diverges there. They
are the analogues of the singular instantons discussed in \cite{hawking:98}.

If $ \alpha < 0$ then there are two types of behaviour. (i) $-1 <
f'''(0) < 0 $. These instantons are compact with an irregular
South pole. (ii) $f'''(0)<-1$. The scale factor of these
instantons increases to a local maximum, decreases to a local minimum,
then has another maximum before decreasing to zero at the South
pole, which is irregular. The instanton therefore has two ``peaks''. 
There is a critical value $\gamma(\alpha)$ such 
that for $\gamma < f'''(0)<-1$ the larger peak is near the North pole
while for $f'''(0)< \gamma$, the larger peak is near the South
pole. It follows that when $f'''(0)=\gamma$ the peaks have the same
size and the instanton is symmetrical about its equator with a regular
South pole.  The scale factor is shown in figure \ref{fig:reg}. 

To summarize, if $\alpha < 0$ then there are two regular compact
instantons, namely the round four sphere and a new ``double bubble'' 
instanton. We shall not have much to say about the new instanton in
this paper since the lack of an analytical solution makes dealing with
perturbations of this instanton rather difficult.

\subsection{Analytic continuation}

The four sphere instanton can be analytically continued to Lorentzian
signature by slicing at the equator $\tilde{\sigma}=\pi/2$ and writing
\be
 \tilde{\sigma} = \frac{\pi}{2} - it/R,
\ee
which yields the metric on a closed de Sitter universe:
\be
 ds^2 = -dt^2 + R^2 \cosh^2 (t/R) d\Omega_3^2.
\ee
The Hubble parameter is $R^{-1}$, which is much smaller than the
Planck mass because $N$ is large. A change of coordinate takes one
from a closed FRW metric to an open FRW metric. 

The double bubble instanton can be analytically continued across
its ``equator'' to give a closed FRW universe. Numerical studies suggest
that this universe rapidly collapses. However, this instanton can also
be continued to an inflationary open universe (the details of the
continuation are the same as in \cite{hawking:98}) and therefore may
give rise to realistic cosmology.

\medskip

\sect{Metric perturbations}

\label{sec:pert}

\subsection{Scalars, vectors and tensors}

\label{subsec:pert}

In this section we shall calculate correlation functions for metric
perturbations around our four sphere instanton. These can then be
analytically continued to yield correlation functions in de Sitter space.
The metric on the perturbed four sphere can be written
\be
 ds^2 = (R^2 \hat{\gamma}_{ij} + h_{ij}) dx^i dx^j,
\ee
where $\hat{\gamma}_{ij}$ denotes the metric on a {\it unit} four sphere.
The perturbation can be decomposed
into scalar, vector and tensor parts with respect to the four sphere:
\be
\label{eqn:decomp}
 h_{ij}(x) = \theta_{ij}(x) + 2 \hat{\nabla}_{(i}
 \chi_{j)}(x) + \hat{\nabla}_i \hat{\nabla}_j \phi(x) +
 \hat{\gamma}_{ij} \psi(x).
\ee
The connection on the unit four sphere is denoted $\hat{\nabla}$. 
$\theta_{ij}$ is a transverse traceless symmetric tensor with respect
to the four sphere:
\be
 \hat{\nabla}_i \theta^{ij} = \theta^i_i = 0,
\ee
where indices $i,j$ are raised and lowered with $\hat{\gamma}_{ij}$.
$\chi_i$ is a transverse vector:
\be
 \hat{\nabla}_i \chi^i = 0.
\ee
There is a small ambiguity in our decomposition - it is invariant
under $\phi \rightarrow \phi+Y$, $\psi \rightarrow \psi + \lambda Y$
where $Y$ satisfies
\be
\label{eqn:gaugeeq}
 \hat{\nabla}_i \hat{\nabla}_j Y + \lambda \hat{\gamma}_{ij} Y = 0.
\ee
This equation can only be solved when  $\lambda=1$. The solutions are
simply the regular $p=1$ spherical harmonics on $S^4$, 
i.e., the regular $p=1$ solutions of
\be
 \left( \hat{\nabla}^2 + p(p+3) \right) Y = 0.
\ee
The spherical harmonics are labelled with integers $p,k,l,m$ with $0
\le |m| \le l \le k \le p$. Hence there are five independent 
spherical harmonics with $p=1$, given in terms of
spherical harmonics $Y_{klm}$ on the three sphere by
\be
\label{eqn:gaugemodes}
 \sin \rho Y_{1lm}, \qquad \cos \rho Y_{000}
\ee
where $\rho$ is the polar angle on the four sphere. These five
harmonics correspond to gauge transformations involving the
five conformal Killing vector fields on the four sphere \cite{gibbons:78}.
If we assume that $\psi$ is regular on $S^4$ then we can expand it in terms of
spherical harmonics. We shall fix the residual gauge ambiguity by
demanding that $\psi$ contain no contribution from the $p=1$ harmonics. 

It is possible to gauge away $\phi$ and $\chi^i$ through a coordinate
transformation on the four sphere of the form $x^i \rightarrow x^i -
\eta^i - \partial^i \eta$, where $\eta^i$ is a transverse vector and
$\eta$ is a scalar. For the moment we shall use a general gauge but
later we will assume that $\phi$ and $\chi^i$ vanish.

\subsection{Matter effective action}

\label{subsec:CFTgen}

We need to calculate the action for metric perturbations. The hardest
part to calculate is the effective action for the matter fields. 
This can be expanded around a round four sphere background:
\ba
 W &=& W^{(0)} - \frac{1}{2} \int d^4 x \sqrt{\gamma}
 \langle T_{ij}(x) \rangle h^{ij}(x) \nonumber \\
 &+& \frac{1}{4} \int d^4 x
 \sqrt{\gamma} \int d^4 x' \sqrt{\gamma}
 h^{ij} (x) \langle T_{ij}(x) T_{kl}(x') \rangle h^{kl}(x') + \ldots
\ea
Here $\gamma$ denotes the determinant of the metric on the sphere.
If we know the one and two point function of the CFT energy
momentum tensor on a round $S^4$ then we can
calculate the effective action to second order in the metric perturbation. The
one point function is given by the conformal anomaly on the round four
sphere. In flat space, the 2-point function is determined entirely by
conformal invariance. 
On the sphere, symmetry determines the 2-point function only
up to a single unknown function \cite{osborn:00}. However, the sphere
is conformally flat so one can calculate the 2-point function on the
sphere using a conformal transformation from flat space. The
energy-momentum tensor transforms anomalously, so there will be a
contribution from the trace anomaly in the transformation. Therefore,
the 2-point function on the sphere is determined by two quantities,
namely the 2-point function in flat space, and the trace anomaly. For
the ${\cal N}=4$ super Yang-Mills theory that we are considering, both
of these quantities are independent of the Yang-Mills coupling. It
follows that the 2-point function on the sphere (or any other
conformally flat space) must be independent of coupling. Therefore the
effective action will be independent of coupling to second order in
the metric perturbation so the effects of strong coupling will not
show up in our results.

For the moment, we shall consider the four sphere to have arbitrary
radius $R$ rather than using the value given by equation
\ref{eqn:Rdef}. Introduce a fictional ball of AdS that has the sphere as its
boundary. Let $\bar{l},\bar{G}$ be the AdS radius and
Newton constant of this region. 
If we take $\bar{l}$ to zero then the sphere is effectively at
infinity in AdS so we can use AdS/CFT to calculate the generating functional
of the CFT
on the sphere. In other words, $\bar{l}$ is acting like a
cut-off in the CFT and taking it to zero corresponds to removing the cut-off.
However the relation
\be
\label{eqn:Ndef2}
 \frac{\bar{l}^3}{\bar{G}} = \frac{2N^2}{\pi},
\ee
implies that if $\bar{l}$ is taken to zero then we must also take
$\bar{G}$ to zero since $N$ is fixed (and large). 

The CFT generating functional is given by evaluating the action of the
bulk metric ${\bf g}$ that matches onto the metric ${\bf
h}$ of the boundary \cite{gubser:98, witten:98}, 
and adding surface counterterms to cancel
divergences as $\bar{l}, \bar{G} \rightarrow 0$ \cite{witten:98,
tseytlin:98, henningson:98, henningson:00, balasubramanian:99,
emparan:99, kraus:99}:  
\be
 W[{\bf h}] = S_{EH}[{\bf g}] + S_{GH}[{\bf g}] + S_1[{\bf h}] +
 S_2[{\bf h}] + S_3[{\bf h}] + S_{ct}[{\bf h}], 
\ee
where $S_{EH}$ denotes the five dimensional Einstein-Hilbert action
with a negative cosmological constant:
\be
 S_{EH} = -\frac{1}{16 \pi \bar{G}} \int d^5 x \sqrt{g} \left(R +
\frac{12}{\bar{l}^2}\right),
\ee 
the overall minus sign arises because we are considering a Euclidean signature
theory. The second term in the action is the Gibbons-Hawking boundary
term \cite{gibbons:77}:
\be
 S_{GH} = -\frac{1}{8 \pi \bar{G}} \int d^4 x \sqrt{h} K,
\ee
where $K$ is the trace of the extrinsic curvature of the
boundary and $h$ the determinant of the induced metric. The first two
surface counterterms are
\be
 S_1 = \frac{3}{8\pi \bar{G} \bar{l}} \int d^4 x \sqrt{h},
\ee
\be
 S_2 = \frac{\bar{l}}{32\pi \bar{G} } \int d^4 x \sqrt{h} R,
\ee
where $R$ now refers to the Ricci scalar of the boundary metric. The
third counterterm is\footnote{
In the prefactor of this equation, $R$ refers to the radius of the
sphere. In the integrand it refers to the Ricci scalar.} 
\be
\label{eqn:CT3div}
 S_3 = -\frac{\bar{l}^3}{64\pi \bar{G}} \left(\log (\bar{l}/R) - \beta
 \right)
 \int d^4 x \sqrt{h} \left(R_{ij} R^{ij} - \frac{1}{3} R^2 \right),
\ee
where $R_{ij}$ is the Ricci tensor of the boundary metric and boundary
indices $i,j$ are raised and lowered with the boundary metric. 
This term is required to cancel logarithmic divergences as
$\bar{l},\bar{G} \rightarrow 0$. The finite part of this term is
arbitrary, which is why we have included the constant $\beta$. The
integrand of this term is a combination of the Euler density and the
square of the Weyl tensor. The former just contributes a constant term
to the action but the latter may have important physical effects so we
shall include it. For a pure gravity theory, adding a Weyl squared
term to the action results in spin-2 ghosts in flat space 
but we shall see that this is not the case when the Yang-Mills theory 
is also included. The final counterterm $S_{ct}$ is the finite $R^2$
counterterm defined in equation \ref{eqn:counterterm}. 

When the four sphere boundary is unperturbed, the metric in the AdS region is
\be
 ds^2 = \bar{l}^2 (dy^2 + \sinh^2 y \hat{\gamma}_{ij} dx^i dx^j),
\ee
and the sphere is at $y=y_0$, where $y_0$ is given 
by $R=\bar{l}\sinh y_0$. Note
that $y_0 \rightarrow \infty$ as $\bar{l} \rightarrow 0$ since $R$ is
fixed. In order to use AdS/CFT for the perturbed sphere, we need to
know how the metric perturbation extends into the bulk. This is done by
solving the Einstein equations linearized about the AdS background. 

Our first task is therefore to solve the Einstein equations 
in the bulk to find the
bulk metric perturbation that approaches $h_{ij}$ on the boundary. 
We shall impose the boundary condition that the metric perturbation
be regular throughout the AdS region. 
The most general perturbation of the bulk metric can be written
\be
 ds^2 = \bar{l}^2 (dy^2 + \sinh^2 y \hat{\gamma}_{ij} dx^i dx^j) + A dy^2 +
 2 B_i dy dx^i + H_{ij} dx^i dx^j. 
\ee
The first step is to decompose the bulk metric fluctuation
into scalar, vector and tensor parts with respect to the four sphere:
\be
\label{eqn:Hdecomp}
 H_{ij}(y,x) = \theta_{ij}(y,x) + 2 \hat{\nabla}_{(i}
 \chi_{j)}(y,x) + \hat{\nabla}_i \hat{\nabla}_j \phi(y,x) +
 \hat{\gamma}_{ij} \psi(y,x).
\ee
The connection on the four sphere is denoted $\hat{\nabla}$. 
$\theta_{ij}$ is a transverse traceless symmetric tensor with respect
to the four sphere:
\be
 \hat{\nabla}_i \theta^{ij} = \theta^i_i = 0,
\ee
where indices $i,j$ are raised and lowered with $\hat{\gamma}_{ij}$.
$\chi_i$ is a transverse vector:
\be
 \hat{\nabla}_i \chi^i = 0.
\ee
We can also decompose $B_i$ into a transverse vector and a scalar:
\be
 B_i = \hat{B}_i + \partial_i B.
\ee
The quantities that we have introduced are gauge dependent. If
we perform an infinitesimal change of coordinate then the five
dimensional metric perturbation undergoes the gauge transformation
\be
 \delta g_{\mu\nu} \rightarrow \delta g_{\mu\nu} + \bar{\nabla}_{\mu}
\xi_{\nu} + \bar{\nabla}_{\nu} \xi_{\mu}.
\ee
We are using Greek letters to denote five dimensional
indices. $\bar{\nabla}$ is the connection with respect to the
background AdS metric. The
gauge parameters $\xi_{\mu}$ can be decomposed with respect to the
four sphere. $\xi_y$ is a scalar and $\xi_i$ can be decomposed into a
transverse vector and a scalar. Thus in total, we have four scalar
degrees of freedom in our metric perturbation but there are two scalar
gauge degrees of freedom so we can only expect two gauge invariant
scalars. Similarly we have two vectors in our metric perturbation, but
one vector gauge degree of freedom so there is only one gauge
invariant vector quantity. The tensor part of the metric perturbation
is gauge invariant. It is easy to check that the following scalar
quantities are gauge invariant:
\be
 \Psi_1 \equiv A - \partial_y \left(\frac{\psi}{\cosh y \sinh y}\right),
\ee
\be
 \Psi_2 \equiv B - \frac{1}{2} \partial_y \phi - \frac{\psi}{2\cosh y
\sinh y} + \coth y \, \phi.
\ee
Note that the residual gauge invariance discussed in section
\ref{subsec:pert} is also present here -- we shall have more to say about
this later on.

The gauge invariant vector quantity is
\be
 X_i \equiv \hat{B}_i - \partial_y \chi_i + 2\coth y \chi_i.
\ee
The gauge invariant tensor is $\theta_{ij}$. 

\subsection{Solving the Einstein equations: scalars and vectors}

The Einstein equation in the bulk is
\be
 R_{\mu\nu} - \frac{1}{2} R g_{\mu \nu} = \frac{6}{\bar{l}^2} g_{\mu\nu}.
\ee
We want to solve this such that our metric matches onto the perturbed
metric on the four sphere boundary. The solution for the unperturbed
sphere is simply AdS. Denote this background metric by
$\bar{g}_{\mu\nu}$. Linearizing around this background yields the
equation
\be
 \bar{\nabla}_{\mu} \bar{\nabla}^{\rho} \delta g_{\rho \nu} +
\bar{\nabla}_{\nu} \bar{\nabla}^{\rho} \delta g_{\rho \mu} -
\bar{\nabla}^2 \delta g_{\mu\nu} - \bar{\nabla}_{\mu}
\bar{\nabla}_{\nu} \delta g^{\rho}_{\rho} = \frac{2}{\bar{l}^2} \delta
 g_{\mu\nu} - \frac{2}{\bar{l}^2} \bar{g}_{\mu\nu} \delta g^{\rho}_{\rho},
\ee
This equation is gauge invariant and can
therefore be expressed in terms of the gauge invariant variables. The
$yy$ component gives
\be
 \hat{\nabla}^2 \Psi_1 - 2 \partial_y \hat{\nabla}^2 \Psi_2 - 4\cosh y
 \sinh y \partial_y \Psi_1 - 8 \sinh^2 y \Psi_1 = 0.
\ee
The vector part of the $iy$ components gives
\be
 \hat{\nabla}^2 X_i = -3 X_i.
\ee
The scalar part of the $iy$ components gives
\be
\label{eqn:iyscalar}
 \partial_i \left(\cosh y \sinh y \Psi_1 - 2\Psi_2\right) = 0.
\ee
The tensor part of the $ij$ components gives
\be
\label{eqn:tensoreq}
 \partial_y^2 \theta_{ij} - 4\coth^2 y \theta_{ij} + \mathrm{cosech}^2 y
\hat{\nabla}^2 \theta_{ij} = 0.
\ee

The vector part of the $ij$ components gives
\be
\label{eqn:ijvec}
 \left( \partial_y + 2\coth y \right) \hat{\nabla}_{(i} X_{j)} = 0.
\ee
The scalar part of the $ij$ components gives
\ba
&{}& \hat{\nabla}_i \hat{\nabla}_j \left( -\Psi_1 + 2\partial_y \Psi_2
 +4\coth y \Psi_2 \right) \nonumber \\ 
 &+& \hat{\gamma}_{ij} \left( \cosh y \sinh y
 \partial_y \Psi_1 + (8\cosh^2 y-2) \Psi_1 + 2\coth y \hat{\nabla}^2
 \Psi_2 \right) = 0.
\ea
Solving equation \ref{eqn:ijvec} yields
\be
 \hat{\nabla}_{(i} X_{j)}(y,x) = \frac{\sinh^2 y_0}{\sinh^2 y}
 \hat{\nabla}_{(i} X_{j)}(y_0,x),
\ee
which is singular at $y=0$. We must therefore take the solution 
\be 
\hat{\nabla}_{(i} X_{j)}(y,x) = 0.
\ee
Thus the gauge invariant vector perturbation vanishes: we are free to
choose $X_i = 0$.

Rearranging the equations for the scalars, one obtains
\be
 \hat{\nabla}^2 \Psi_1 = -4 \Psi_1
\ee
and
\be 
 \left( \cosh y \sinh y \partial_y  + (4\cosh^2 y -2)\right) 
 \left( \hat{\nabla}_i \hat{\nabla}_j + \hat{\gamma}_{ij} \right)
 \Psi_1 = 0.
\ee
This has the solution
\be
 \left( \hat{\nabla}_i \hat{\nabla}_j + \hat{\gamma}_{ij} \right)
 \Psi_1(y,x) = \frac{\sinh^2 y_0 \cosh^2 y_0}{\sinh^2 y\cosh^2 y}
 \left( \hat{\nabla}_i \hat{\nabla}_j + \hat{\gamma}_{ij} \right)
 \Psi_1 (y_0,x).     
\ee
Once again, this is singular at $y=0$ unless we take
\be
 \left( \hat{\nabla}_i \hat{\nabla}_j + \hat{\gamma}_{ij} \right)
 \Psi_1(y,x) = 0.
\ee
There is a regular solution to this equation, however it is simply an
artifact of the ambiguity in our metric decomposition
discussed in section \ref{subsec:pert} (see equation \ref{eqn:gaugeeq})
so $\Psi_1$ can be consistently set to zero.
Equation \ref{eqn:iyscalar} then implies that $\Psi_2$ is
an arbitrary function of $y$. This is again related to an ambiguity in
the metric decomposition: we are free to add
an arbitrary function of $y$ to $\phi$ without changing the metric
perturbation. Hence we can choose $\Psi_2 = 0$.

To summarize: we have solved the bulk Einstein equation for the gauge
invariant vector and scalars, obtaining the result 
\be
\label{eqn:soln}
 \Psi_1 = \Psi_2 = X_i = 0.
\ee

So far we have been working in a general gauge. We shall now specialize
to Gaussian normal coordinates, in which we define $ly$ to be the
geodesic distance from some origin in our ball of perturbed AdS, and then
introduce coordinates $x^i$ on surfaces of constant $y$ (which have spherical
topology). In these coordinates we have
\be
 A = B = \hat{B}_i = 0.
\ee
The presence of a metric perturbation implies that the boundary of the
ball is not at constant geodesic distance from the origin. 
Instead it will be at a position 
\be
 y = y_0 + \xi(x).
\ee
We can now use our solution \ref{eqn:soln} to write down the bulk
metric perturbation in Gaussian normal coordinates:
\be
 \psi(y,x) = f(x) \sinh y \cosh y,
\ee
\be
 \phi(y,x) = f(x) \sinh y \cosh y + g(x) \sinh^2 y,
\ee
\be
 \chi_i(y,x) = \hat{\chi}_i (x) \sinh^2 y,
\ee
where $f$, $g$ are arbitrary functions of $x$ and $\hat{\chi}_i$ is an
arbitrary transverse vector function of $x$. We now appear to have
three independent scalar functions of $x$ to deal with (namely $f$,
$g$ and $\xi$). These should be specified by demanding that the bulk
metric perturbation match onto the boundary metric
perturbation. However the boundary metric perturbation is specified by
only two scalars. We therefore need another boundary condition:
regularity at the origin. Solutions proportional to $\sinh y \cosh y$
are unacceptable since they lead to
\be
 \bar{g}^{\mu\nu} \delta g_{\mu\nu} \propto \coth y,
\ee
which is singular at $y=0$. We must therefore set $f(x) = 0$. 
To first order, the induced metric perturbation on the boundary
is
\be
 h_{ij}(x) = H_{ij}(y_0,x) + 2l^2 \sinh y_0 \cosh y_0 \hat{\gamma}_{ij} \xi.
\ee
Recall that $H_{ij}$ is given by equation \ref{eqn:Hdecomp}.
The left hand side is 
decomposed into scalar, vector and tensor pieces in equation
\ref{eqn:decomp}\footnote{
We apologize for our slightly confusing notation:
$\psi(x)$, $\phi(x)$ and $\chi_i(x)$ in equation \ref{eqn:decomp}
have, so far, nothing to do with the bulk quantities 
$\psi(y,x)$, $\phi(y,x)$ and $\chi_i(y,x)$.}. We can substitute the
solution for the bulk metric perturbation into the right hand side and
read off 
\be
 \psi(x) = 2l^2 \sinh y_0 \cosh y_0 \, \xi(x),
\ee
\be
 \phi(x) = g(x) \sinh^2 y_0,
\ee
\be
 \chi_i(y,x) = \hat{\chi}_i (x) \sinh^2 y.
\ee
These equations determine $\xi(x)$, $g(x)$ and $\hat{\chi}_i(x)$ in
terms of the boundary metric perturbation. In section \ref{subsec:pert},
we showed that $\phi(x)$ and $\chi_i(x)$ could be gauged away so we
shall now set 
\be
 g(x) = 0, \qquad \hat{\chi}_i(x) = 0.
\ee
This implies that
\be
 \phi(y,x) = \psi(y,x) = 0, \qquad \chi_i (y,x) = 0.
\ee
In other words, all scalar and vector perturbations vanish in the
bulk: the bulk perturbation is transverse and traceless. The only
degrees of freedom that remain are therefore the bulk tensor
perturbation and the scalar perturbation $\xi(x)$ describing the
displacement of the boundary.

\subsection{Tensor perturbations}

The tensor perturbations are less trivial: we have to solve equation
\ref{eqn:tensoreq}. This was done in \cite{hawking:00} by expanding in
tensor spherical harmonics $H_{ij}^{(p)}$. These obey
\be
 \hat{\gamma}^{ij} H^{(p)}_{ij}(x) = \hat{\nabla}^i H^{(p)}_{ij}(x) = 0,
\ee
and they are regular tensor eigenfunctions of the Laplacian:
\be
 \hat{\nabla}^2 H^{(p)}_{ij} = \left(2-p(p+3)\right) H^{(p)}_{ij},
\ee
where $p=2,3,\ldots$. We have suppressed extra labels $k,l,m,\ldots$ on
these harmonics. The harmonics are orthonormal with respect to the
obvious inner product. Further properties are given in
\cite{higuchi:87}. 

The boundary condition at $y=y_0$ is\footnote{
The boundary is actually at $y=y_0 + \xi(x)$, which gives higher order
corrections. These would appear at third order in the action as
couplings between tensors and scalars.} $\theta_{ij}(y_0,x) =
\theta_{ij}(x)$, where $\theta_{ij}(x)$ is the tensor part of the
metric perturbation on the boundary. Imposing this condition together
with regularity at the origin gives a unique bulk solution \cite{hawking:00}
\be
\label{eqn:metricsol}
 \theta_{ij}(y,x) = \sum_p \frac{f_p(y)}{f_p(y_0)} H^{(p)}_{ij}(x) \int d^4x'
 \sqrt{\hat{\gamma}} \theta^{kl}(x') H^{(p)}_{kl}(x'),
\ee
where $f_p$ is given in terms of a hypergeometric function:
\be
\label{eqn:fdef}
 f_p (y) = \frac{\sinh^{p+2} y}{\cosh^p y} 
 {}_2F_1(p/2,(p+1)/2, p+5/2, \tanh^2 y).
\ee

\subsection{The gravitational action}

We have now solved the Einstein equations in the bulk and found a
solution that matches onto the metric perturbation of the
boundary. The next step is to compute the action of this solution. 
The bulk contribution from the Einstein-Hilbert action with
cosmological constant is
\ba
\label{eqn:EH1}
 S_{bulk} &=& \frac{\bar{l}^3}{2\pi \bar{G}} \int d^4 x \sqrt{\hat{\gamma}}
 \int_0^{y_0+\xi} dy \sinh^4 y \\
 &-&\frac{1}{16\pi \bar{G}} 
 \int d^5 x \sqrt{\bar{g}} \left[ -\left(\bar{R}_{\mu\nu}
 - \frac{1}{2} \bar{R} \bar{g}_{\mu\nu} -\frac{6}{l^2} \bar{g}_{\mu\nu}
 \right) \delta g^{\mu\nu} - \delta g_{\mu\nu}
 \Delta_L^{\mu\nu\rho\sigma} \delta g_{\rho\sigma} \right]. \nonumber
\ea
The term that is first order in $\delta g_{\mu\nu}$ 
will vanish because the background obeys the
Einstein equation. The second order term involves the Lichnerowicz
operator (generalized to include the effect of a cosmological
constant) $\Delta_L$, which is a second order differential operator
with the symmetry property
\be
 \Delta_L^{\mu\nu\rho\sigma} = \Delta_L^{\rho\sigma\mu\nu}.
\ee
This term vanishes because the perturbation is on shell, i.e.,
\be
 \Delta_L^{\mu\nu\rho\sigma} \delta g_{\rho\sigma} = 0.
\ee
We are left simply with the background contribution 
\ba
 S_{bulk} &=& \frac{\bar{l}^3}{2\pi \bar{G}} \int d^4 x \sqrt{\hat{\gamma}} 
 \int_0^{y_0+\xi} dy \sinh^4 y \\
 &=& \frac{\bar{l}^3 \Omega_4}{2\pi \bar{G}} \int_0^{y_0} dy \sinh^4 y +
 \frac{\bar{l}^3}{8\pi \bar{G}} 
 \int d^4 x \sqrt{\hat{\gamma}} \left( 4 \sinh^4
 y_0 \, \xi + 8 \sinh^3 y_0 \cosh y_0 \, \xi^2 \right), \nonumber
\ea
where $\Omega_4$ denotes the volume of a unit four sphere.
Of course, in order to rearrange the Einstein-Hilbert action into the
form \ref{eqn:EH1} we have to integrate by parts several times, giving
rise to surface terms. These will depend on derivatives of the bulk 
metric perturbation evaluated at the boundary. Since there are only 
tensor degrees of freedom excited in the bulk, only tensors will 
occur in these surface terms -- there will be no dependence on
$\xi$. The surface terms are
\be
 S_{surf} = \frac{\bar{l}^3}{16\pi \bar{G}} \int d^4 x \sqrt{\hat{\gamma}}
\left( \frac{3}{4\bar{l}^4} \theta^{ij} \partial_y \theta_{ij} - \frac{\coth
y_0}{\bar{l}^4}  \theta^{ij} \theta_{ij} \right).
\ee

The second contribution to the gravitational action is the
Gibbons-Hawking term. In evaluating this, it is important to remember
that the unit normal to the boundary changes when we perturb the bulk
metric. The boundary is a hypersurface defined by the condition
$f(y,x) \equiv y-\xi(x) =y_0$. The unit normal is therefore given to
second order by
\be
 n = \bar{l} \left(1 - \frac{\partial_i \xi \partial^i \xi}{2 \sinh^2 y}
 \right) dy - \bar{l} \partial_i \xi dx^i.
\ee
Note that this holds for a range of $y$ and therefore defines a unit
covector field that is normal to the family of hypersurfaces
$f={\rm constant}$. In other words, it defines an extension of the
unit normal on the boundary into a neighbourhood of the boundary. 
Written as a vector, the normal takes the form
\be
 n = \frac{1}{\bar{l}}  
 \left(1 - \frac{\partial_i \xi \partial^i \xi}{2 \sinh^2 y}
 \right) \frac{\partial}{\partial y} - \left(\frac{\partial^i
 \xi}{\bar{l} \sinh^2 y} 
 - \frac{\theta^{ij}(y,x) \partial_j \xi}{\bar{l}^3 \sinh^4 y} \right)
 \frac{\partial}{\partial x^i}, 
\ee
where $\theta_{ij}(y,x)$ is the bulk tensor perturbation.
The trace of the extrinsic curvature is
\be
 K \equiv \nabla_{\mu} n^{\mu}.
\ee
In evaluating this one must take account of both the perturbation in the
unit normal and the perturbation in the connection. The result is
\ba
 K &=& \frac{4}{\bar{l}} \coth y - \frac{1}{\bar{l} \sinh^2 y} 
 \hat{\nabla}^2 \xi -
\frac{\cosh y}{\bar{l} \sinh^3 y} \partial_i \xi \partial^i \xi \nonumber \\
 &+& \frac{1}{\bar{l}^3 \sinh^4 y} \theta^{ij} \hat{\nabla}_i \hat{\nabla}_j
 \xi - \frac{1}{2 \bar{l}^5 \sinh^4 y} \theta^{ij} \partial_y \theta_{ij} +
 \frac{\cosh y}{\bar{l}^5 \sinh^5 y} \theta^{ij} \theta_{ij}.
\ea
This has to be evaluated at $y=y_0 + \xi$. To evaluate $\sqrt{\gamma}$
on the boundary, we need to know
the induced boundary metric perturbation to {\it second} order:
\be
\label{eqn:metricpert}
 h_{ij}(x) = \theta_{ij}(y_0,x) + 2\bar{l}^2 \sinh y_0 \cosh y_0
 \hat{\gamma}_{ij} \xi + \bar{l}^2 
 \left(2\sinh^2 y_0 + 1 \right) \hat{\gamma}_{ij}
 \xi^2 + \bar{l}^2 \partial_i \xi \partial_j \xi + \xi \partial_y
 \theta_{ij}.
\ee
These results can now be substituted into the Gibbons-Hawking term,
yielding
\ba
 S_{GH} &=& -\frac{\bar{l}^3}{8\pi \bar{G}} 
 \int d^4 x \sqrt{\hat{\gamma}} \left[ 4
 \cosh y_0 \sinh^3 y_0 + \sinh^2 y_0 \left(16 \sinh^2 y_0 + 12 \right)
 \xi \right. \\ &+& \left. \cosh y_0 \sinh y_0 \left( 32
 \sinh^2 y_0 + 12 \right) \xi^2 - 3\cosh y_0 \sinh y_0 \, \xi
 \hat{\nabla}^2 \xi -\frac{1}{2 \bar{l}^4} \theta^{ij} \partial_y
 \theta_{ij} \right]. \nonumber
\ea
We have integrated some terms by parts. So far, we have expressed the
scalar part of the action in terms of $\xi$. However, we really want
to express everything in terms of the induced metric on the boundary,
which has scalar part $\psi(x)$. This can be done by taking the trace
of equation \ref{eqn:metricpert} and solving for $\xi$ in terms of
$\psi$ to second order, giving
\be
 \xi = \frac{\psi}{2 \bar{l}^2 \sinh y_0 \cosh y_0} - \frac{\left(2 \sinh^2
 y_0 +1 \right) \psi^2} {8 \bar{l}^4 \sinh^3 y_0 \cosh^3 y_0} -
 \frac{\partial_i \psi \partial^i \psi}{32 \bar{l}^4 \sinh^3 y_0 \cosh^3 y_0}.
\ee
The total contribution from the Einstein-Hilbert and Gibbons-Hawking
terms is given by the sum of the following
\be
\label{eqn:Sgrav0}
 S_{grav}^{(0)} = -\frac{3 \bar{l}^3 \Omega_4}{2\pi \bar{G}} \int_0^{y_0} dy
 \sinh^2 y \cosh^2 y,
\ee
\be
\label{eqn:Sgrav1}
 S_{grav}^{(1)} = -\frac{3 \bar{l}^3}{4 \pi \bar{G}} \int d^4 x
 \sqrt{\hat{\gamma}} \frac{1}{\bar{l}^2} \cosh y_0 \sinh y_0 \, \psi, 
\ee
\ba
\label{eqn:Sgrav2}
 S_{grav}^{(2)} = -\frac{\bar{l}^3}{8\pi \bar{G}} 
 \int d^4 x \sqrt{\hat{\gamma}}
 \left[ \frac{3 \left(2 \sinh^2 y_0 + 1\right) \psi^2}{2 \bar{l}^4 \sinh y_0
 \cosh y_0 } - \frac{3 \psi \hat{\nabla}^2 \psi}{8 \bar{l}^4 \sinh y_0 \cosh
 y_0} \right. \nonumber \\ \left. -\frac{1}{8\bar{l}^4} \theta^{ij}
 \partial_y \theta_{ij} - \frac{\coth y_0}{2 \bar{l}^4} \theta^{ij}
 \theta_{ij} \right].
\ea
We can now expand
the action in powers of $\bar{l}/R$ (using $\sinh y_0 = R/\bar{l}$). 
This gives terms that diverge as $\bar{l}^{-4}$ and $\bar{l}^{-2}$ 
as $\bar{l}$ goes to zero. For the scalar perturbation, 
these divergences are cancelled by the counter terms $S_1$
and $S_2$. For the tensor perturbation (dealt with in
\cite{hawking:00}), the third counter term $S_3$ is needed
to cancel a logarithmic divergence\footnote{
This counter term is formed from the Euler number and the square of
the Weyl tensor, neither of which is affected by scalar
perturbations. $S_3$ therefore does not contribute to the action for
scalar perturbations.}.

The final term that we have to include in the effective action
is the finite counter term $S_{ct}$. Evaluating this to second order gives
\be
 S_{ct} = \frac{3\alpha N^2 \Omega_4}{4\pi^2} 
 + \frac{3 \alpha N^2 }{64 \pi^2 R^4} \int d^4 x \sqrt{\hat{\gamma}}
\left( \psi \hat{\nabla}^4 \psi + 4\psi \hat{\nabla}^2 \psi +
\frac{2}{3} \theta^{ij} \hat{\nabla}^2 \theta_{ij} - \frac{4}{3}
\theta^{ij} \theta_{ij} \right).
\ee
The final result for the Yang-Mills effective action is
\be
\label{eqn:CFTgenfun}
 W = W^{(0)} + W^{(1)} + W^{(2)} + \ldots
\ee
where 
\be
 W^{(0)} = -\frac{3 \beta N^2 \Omega_4}{8 \pi^2} + \frac{3\alpha N^2
 \Omega_4}{4\pi^2} + \frac{3N^2 \Omega_4}{32 \pi^2} \left( 4\log 2 -1 \right), 
\ee
\be
\label{eqn:CFTgenfun1}
 W^{(1)} = \frac{3N^2}{16 \pi^2 R^2} \int d^4 x \sqrt{\hat{\gamma}} \, \psi,
\ee
\ba
\label{eqn:CFTgenfun2}
 W^{(2)} &=& -\frac{3 N^2}{64 \pi^2 R^4} \int d^4 x \sqrt{\hat{\gamma}}
\left[ \psi \left(\hat{\nabla}^2 +2\right) \psi - \alpha \psi
\left(\hat{\nabla}^4 + 4 \hat{\nabla}^2 \right) \psi \right] 
 \nonumber \\
 &+&  \frac{N^2}{256 \pi^2 R^4} \sum_p \left(\int d^4x'
 \sqrt{\hat{\gamma}} \, \theta^{ij}(x') H^{(p)}_{ij}(x') \right)^2 \\
 &\times &  \left( \Psi(p) + 2\beta p(p+1)(p+2)(p+3) -4\alpha p(p+3) 
 \right) \nonumber,
\ea
where
\ba
 \Psi(p) & = &  p(p+1)(p+2)(p+3) \left[\psi(p/2+5/2) + \psi(p/2+2) -
 \psi(2) - \psi (1)\right]\nonumber\\
& &  + p^4+2p^3-5p^2-10p -6.   
\ea 
The scalar perturbations have an action that can be expressed simply
in position space. However, the tensor perturbations are given in
momentum space where they have an action with complicated non-polynomial
dependence on $p$. This corresponds to a non-local action in position
space. At large $p$ it behaves like $p^4 \log p$, as expected from
the flat space result for $\langle T_{ij}(x) T_{i'j'}(x') \rangle$
\cite{gubser:98}. 

\subsection{Metric correlation functions}

Our theory is just four dimensional Einstein gravity coupled to the
Yang-Mills theory, with action
\be
 S = -\frac{1}{16 \pi G} \int d^4 x \sqrt{g} R + W,
\ee
where we have not included a Gibbons-Hawking term because the
instanton has no boundary. 
Note that we are still working in Euclidean signature.
$W$ denotes the Yang-Mills effective action,
including the effect of the finite counterterms. $G$ is the four
dimensional Newton constant. In order to compute
the two point correlation functions of metric perturbations we need to
calculate the terms in $S$ that are quadratic in the metric
perturbations described by $\theta_{ij}$ and $\psi$.

To second order, the Einstein-Hilbert action of the perturbed four
sphere is
\be
\label{eqn:EHaction}
 S_{EH} = -\frac{3 \Omega_4 R^2}{4 \pi G} - \frac{3}{4 \pi G}
 \int d^4 x \sqrt{\hat{\gamma}} \psi +
 \frac{1}{16 \pi G R^2} \int d^4 x
 \sqrt{\hat{\gamma}} \left( \frac{3}{2} \psi \hat{\nabla}^2 \psi 
 + 2 \theta^{ij} \theta_{ij} - \frac{1}{4} \theta^{ij} \hat{\nabla}^2
 \theta_{ij} \right).
\ee
Adding the Yang-Mills effective actions gives the total action. This
has a non-vanishing piece linear in $\psi$. Varying $\psi$ fixes $R$
to take the value given by equation \ref{eqn:Rdef}, which implies that
the linear term vanishes. Equation \ref{eqn:Rdef} can be used to write
$G$ in terms of $R$, which brings the quadratic part of the scalar action to
the form\footnote{
If $\alpha = 0$ then this is almost exactly the same as the scalar action one
would obtain for perturbations about a de Sitter solution supported by
a cosmological constant. The only difference is that the overall sign is
reversed. This implies that, with the exception of the homogeneous
mode, the conformal factor problem of Euclidean quantum gravity is
solved by coupling to the Yang-Mills theory when $\alpha = 0$.}
\be
 S_{scalar} = \frac{3 N^2}{128 \pi^2 R^4} \int d^4 x
 \sqrt{\hat{\gamma}} \, \psi \left( 2\alpha \hat{\nabla}^2 - 1 \right) 
 \left( \hat{\nabla}^2 + 4 \right)
 \psi,
\ee
and the quadratic part of the tensor action becomes
\be
 S_{tensor} = \frac{N^2}{256 \pi^2 R^4} \sum_p \left(\int d^4x'
 \sqrt{\hat{\gamma}} \, \theta^{ij}(x') H^{(p)}_{ij}(x') \right)^2
 F(p,\alpha,\beta), 
\ee 
where
\be
\label{eqn:Fdef}
 F(p,\alpha,\beta) = p^2 + 3p +6 + \Psi(p) + 
 2 \beta p(p+1)(p+2)(p+3) - 4\alpha p(p+3). 
\ee
From these expressions we can read off the correlation functions of
metric perturbations:
\be
\label{eqn:scalarcorrelator}
 \langle \psi (x) \psi (x') \rangle = \frac{32 \pi^2 R^4}{3N^2
 (-\alpha) (4+m^2)} \left[ \frac{1}{-\hat{\nabla}^2 + m^2} -
 \frac{1}{-\hat{\nabla}^2 -4} \right],
\ee
where
\be
 m^2 = \frac{1}{2\alpha}.
\ee
The tensor correlator is
\be
\label{eqn:tensorcorrelator}
 \langle \theta_{ij}(x) \theta_{i'j'}(x') \rangle = \frac{128 \pi^2
 R^4}{N^2} \sum_{p=2}^{\infty}
 W^{(p)}_{ij i'j'}(x,x') F(p,\alpha,\beta)^{-1},
\ee
where the bitensor $W^{(p)}_{iji'j'}(x,x')$ is defined as
\be
 W^{(p)}_{iji'j'}(x,x') = \sum_{k,l,m,\ldots} H^{(p)}_{ij}(x)
 H^{(p)}_{i'j'}(x'),
\ee
with the sum running over all the suppressed labels $k,l,m,\ldots$ of the
tensor harmonics on the four sphere.

\sect{Analytic structure of propagators}

\label{sec:analytic}

\subsection{Flat space limit}

Before analyzing our correlation functions we shall consider the
analagous functions in flat space. This will allow us to constrain the
allowed values of the parameters $\alpha$ and $\beta$, which will be
important when we return to the de Sitter case. 

Recall that in equations \ref{eqn:CFTgenfun1}, \ref{eqn:CFTgenfun2} 
and \ref{eqn:EHaction}, the radius $R$ is arbitrary. To avoid
confusion, we shall now denote this arbitrary radius by $\tilde{R}$ to
distinguish it from the on-shell value $R$, given by equation
\ref{eqn:Rdef}. We can recover flat space
results by taking $\tilde{R} \rightarrow \infty$. Before taking this limit, 
we first replace the dimensionless momentum $p$ with the dimensionful
momentum $k = p/\tilde{R}$. 

There is no conformal anomaly in flat space and the scalar
$\psi$ corresponds to a conformal transformation. Therefore, the only
matter contribution to the scalar propagator comes from the term in
the Yang-Mills action that breaks the conformal invariance, 
namely the finite counter
term $S_{ct}$. The other contribution to the scalar correlator comes
from the Einstein-Hilbert action. One obtains
\be
 \langle \psi(x) \psi(x') \rangle \propto \frac{1}{-\partial^2 + M^2} - 
 \frac{1}{-\partial^2},
\ee
with a positive constant of proportionality. $M^2$ is given by
\be
\label{eqn:flatscalarmass}
 M^2 = -\frac{1}{\alpha R^2},
\ee
where $R$ is given by equation \ref{eqn:Rdef}, although we emphasize
that we are now working in flat space.
The second term in the propagator describes a massless scalar ghost. 
This can be dealt with by gauge fixing the
action. The first term is more worrying. If $\alpha>0$ then it
describes a tachyon. We regard this as undesirable: we do not want
flat space to be an unstable solution of our theory. We shall
therefore always take $\alpha < 0$, which gives a massive scalar in
flat space.

For the tensor propagator, the limit $\tilde{R} \rightarrow \infty$ makes the
coefficient of the third counterterm $S_3$ diverge. To cancel this
divergence, introduce a length scale $\rho$ defined by
\be
 \beta = \log (\rho/\tilde{R}).
\ee
The $\tilde{R}$ dependence in the coefficient of the third counterterm then
drops out, leaving a finite coefficient depending on the
renormalization scale $\rho$. The $\tilde{R} \rightarrow \infty$ limit of the
propagator is similarly well-defined. The result is proportional to
\be
\label{eqn:tomboulis}
 \frac{1}{k^2 \left\{1+R^2 k^2 \left[1 + \log(k^2 \rho^2/4) \right]
 \right\}},
\ee
Our propagator is of exactly the same form as given by Tomboulis
\cite{tomboulis:77} in his analysis of the effects of large $N$ matter
on the flat space graviton propagator. The propagator is defined for $k^2 >
0$. It can be analytically
continued into the complex $k^2$ plane by taking a branch cut for the
logarithm along the negative real axis. There are generally two poles
present, with positions dependent on $\rho$. If $\rho < 2R/e$ then
these poles are on the positive real axis. One has
positive residue and the other negative residue, so they correspond to
a tachyon and a ghost. 
As $\rho \rightarrow 2R/e$, the two poles move together
and merge to form a double pole. For $\rho > 2R/e$, this double pole
splits into a pair of complex conjugate poles which move off into the
complex $k^2$ plane. The modulus $r$ and phase $\theta$ of $k^2$ 
at these poles are related by
\be
 r = \frac{\sin \theta}{R^2 \theta}.
\ee
$\theta$ is given by solving
\be
 \theta \cot \theta = - \left(1 + \log \frac{\rho^2}{4R^2} + \log
 \frac{\sin \theta}{\theta} \right),
\ee
which is straightforward to analyze graphically. The solution obeys
$\theta \rightarrow \pm \pi$ and $r \rightarrow 0$ as 
$\rho \rightarrow \infty$. 

The presence of tachyons for small $\rho$ was not mentioned by Tomboulis
since he implicitly assumed $\rho \gg R$. Since we want flat space to be
a stable solution of our theory, we shall take $\rho > 2R/e$ 
when we consider the propagator in de Sitter space. This corresponds
to taking $\beta > \log 2 - 1$.

It is interesting to note that changing $\rho$ changes the coefficient
of the third counter term $S_3$ by a finite amount. This corresponds
to introducing a finite counter term involving the Euler number and
the square of the Weyl tensor. The former is left unchanged by metric
perturbations. However, the latter is known to give rise to spin-2 ghosts in
a pure gravity theory. Such ghosts do not appear in our model:
coupling to the CFT removes them.

\subsection{Scalar propagator on the sphere}

\label{subsec:scalarsphere}

Equation \ref{eqn:scalarcorrelator} is the propagator of scalar metric
perturbations on a spherical instanton supported by the conformal
anomaly of the CFT. 
The first term in the propagator describes a particle with physical
mass-squared $m^2/R^2 =(2 \alpha R^2)^{-1}$. Since we are assuming $\alpha <
0$, we have $m^2 < 0$ so this particle is a tachyon. This is good
because we do not want the spherical solution to be stable since that
would lead to a Lorentzian de Sitter solution in which inflation never
ends. 
Making $\alpha$ more negative makes the tachyon mass squared less negative,
and therefore makes the instability weaker. This suggests that if
$\alpha$ is sufficiently negative then inflation will last for a long
time. We shall make this more precise later. 

The second term in the propagator describes a ghost.
This is the normal scalar mode of gravity that is canceled by the
scalar parts of the Fadeev-Popov ghosts \cite{gibbons:78}. These
ghosts supply a determinant that cancels the $(\hat{\nabla}^2 + 4)$
factor in the scalar action. The propagator can then be read off from
the action:
\be
\label{eqn:scalarnoghost}
 \langle \psi(x) \psi(x') \rangle = \frac{32 \pi^2 R^4}{3|\alpha | N^2}
\left( -\hat{\nabla}^2 + m^2 \right)^{-1}.
\ee
This propagator can be written in momentum space as
\be \label{eqn:scor}
 \langle \psi(x) \psi(x') \rangle = \frac{32 \pi^2 R^4}{3|\alpha | N^2}
\sum_{p=0}^{\infty} \frac{W^{(p)}(\mu(x,x'))}{p(p+3)+m^2},
\ee
where the biscalar $W^{(p)}$ is a function of the geodesic distance
$\mu$ between $x$ and $x'$, given by
\be
 W^{(p)}(\mu(x,x')) = \sum_{k,l,m} H^{(p)}(x) H^{(p)}(x'),
\ee
where $H^{(p)}$ denote spherical harmonics on the four sphere and the
sum runs over the suppressed eigenvalues $k,l,m$. 

Notice that there are many negative modes 
if $\alpha $ is negative and close to zero. However, if $\alpha <
-1/8$ then only the homogenous ($p=0$) negative mode remains.
To compute the primordial density fluctuations in the 
microwave background radiation we are interested in the two-point
function with the homogenous mode projected out \cite{hawking:83}.
Notice also that the Fadeev-Popov ghosts fix the residual gauge
ambiguity associated with the $p=1$ modes. These modes no longer have
zero action and therefore cannot be regarded as gauge.

\subsection{Tensor propagator on the sphere}

\label{subsec:tensorprop}

The tensor propagator (equation \ref{eqn:tensorcorrelator}) 
has an interesting analytic structure. The momentum space 
propagator is proportional to $F(p,\alpha,\beta)^{-1}$, where $F$ is
given by equation \ref{eqn:Fdef}.

For a physical interpretation, we need to study the behaviour of $F$ in the
complex $\lambda_p$ plane, where
$\lambda_p = p(p+3)-2$ is the eigenvalue of $-\hat{\nabla}^2$. We must
therefore first write the propagator as a function of $\lambda_p$. Since
\be
 p = -\frac{3}{2} \pm \sqrt{\frac{17}{4} + \lambda_p},
\ee
we must choose a branch for the square root. The Euclidean propagator
is defined as a sum over $p=2,3,\ldots$, for which $\lambda_p$ is
positive. We must therefore take the positive sign for the square root.
The analytic continuation into the complex $\lambda_p$ plane is given by
taking a branch cut along the negative axis for $\lambda_p <
-17/4$. $p$ has positive imaginary part just above the cut and
negative imaginary part just below the cut. Note that $\mathrm{Re}(p)
\ge -3/2$. The branch cut corresponds
to a continuum of multi-particle states. The imaginary part of the
propagator is discontinuous across the cut. In general, the absence of
negative norm states implies that the imaginary part of the propagator
just below the cut minus the imaginary part just above the cut should 
be positive, which is indeed the case for our tensor propagator.

\begin{figure}
\begin{picture}(0,0)
\put(12,0){\tiny{$-1.5$}}
\put(65,0){\tiny{$-1.0$}}
\put(110,0){\tiny{$-0.5$}}
\put(162,0){\tiny{$0.0$}}
\put(212,0){\tiny{$0.5$}}
\put(233,0){\tiny{$-0.1$}}
\put(335,0){\tiny{$0.0$}}
\put(430,0){\tiny{$0.1$}}
\put(10,25){\tiny{$0.0$}}
\put(10,67){\tiny{$2.0$}}
\put(10,108){\tiny{$4.0$}}
\put(10,148){\tiny{$6.0$}}
\put(230,28){\tiny{$0.0$}}
\put(230,76){\tiny{$0.1$}}
\put(230,125){\tiny{$0.2$}}
\end{picture}
\centering{\psfig{file=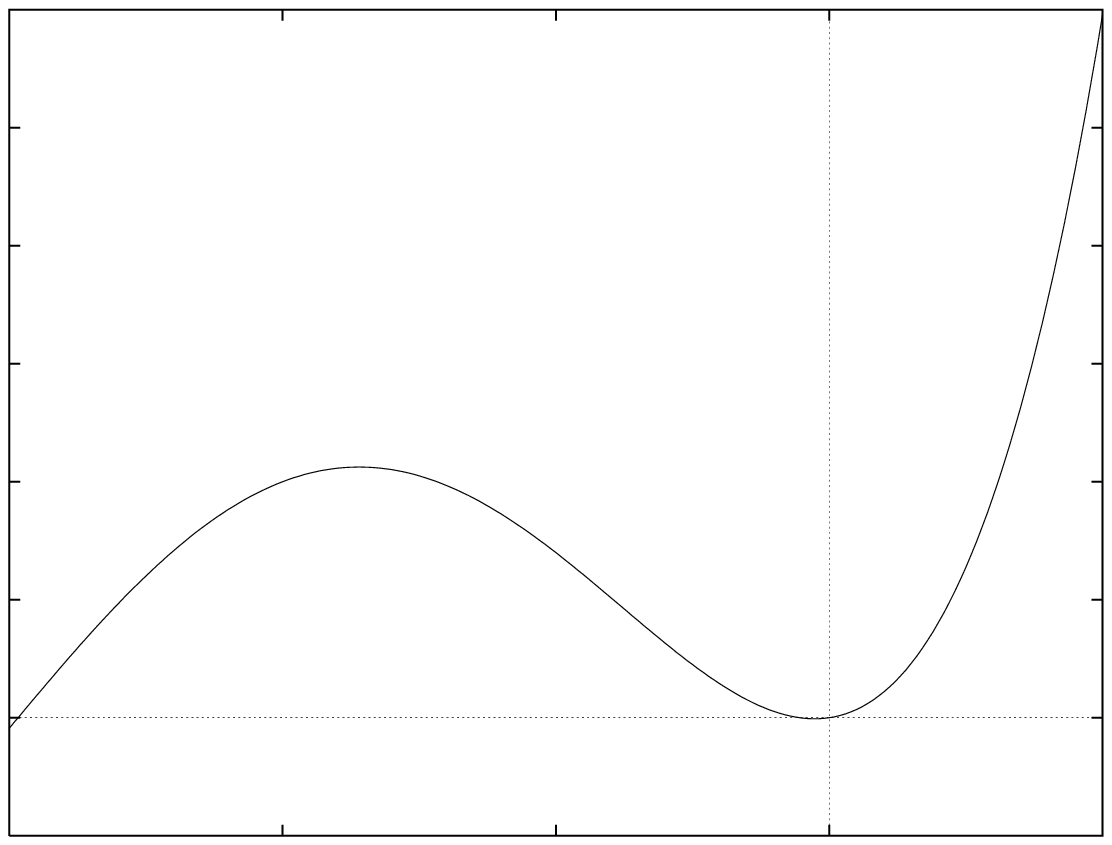,width=3.in}
 \psfig{file=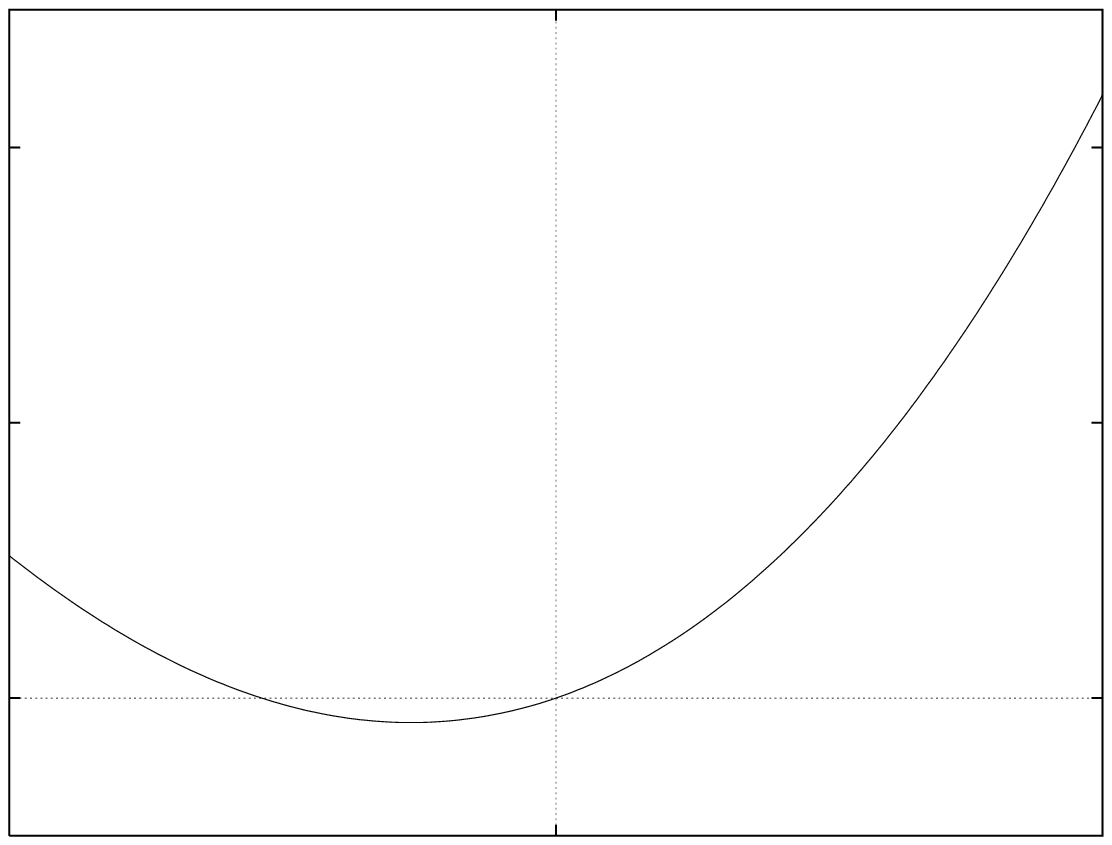,width=3.in}}
\caption{Inverse propagator $F(p,0,0)$ for $-3/2 \le p \le 1/2$ and
 $-0.1 < p < 0.1$. The graph grows monotonically for $p>0$. 
 There are zeroes at $p \approx -1.48$ (massive particle), 
 $p \approx -0.054$ (ghost) and $p=0$ (massless graviton).}
\label{fig:FR0}
\end{figure}

It is also possible for the tensor propagator to have discrete poles
in the $\lambda_p$ plane. Poles on the real axis are of particular
importance. If such a pole occurs at positive $\lambda_p$ then it
corresponds to a tachyon. In fact, since the graviton in de Sitter
space has an equation of motion with $\lambda_p = -2$, it seems appropriate to
regard particles with $\lambda_p > -2$ as tachyons. 
If a pole on the real axis has negative residue then it corresponds 
to a ghost.  

Our propagator always has a pole at $\lambda_p = -2$ ($p=0$), corresponding to
the massless graviton in de Sitter space. Support for this
interpretation comes from observing that transverse traceless tensor
harmonics have 5 degrees of freedom. However, the mode with $p=0$
mixes with transverse vector harmonics, which have 3 degrees of
freedom. Thus the $p=0$ mode has 3 gauge degrees of freedom, leaving 2
physical degrees of freedom, as appropriate for a massless spin-2 particle. 

\begin{figure}
\centering{\psfig{file=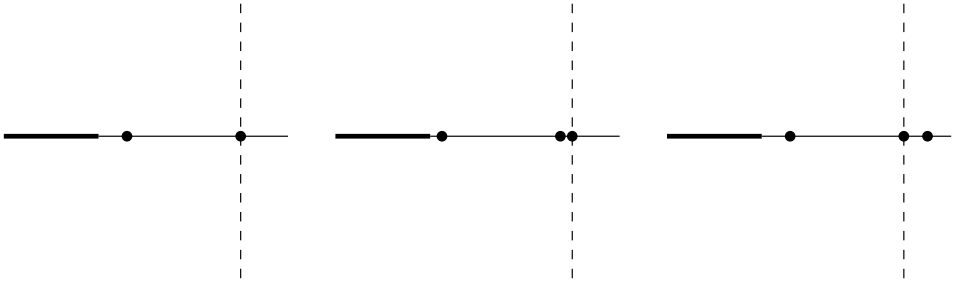,width=5.in}}
\caption{Analytic structure of the tensor propagator in the
complex $\lambda_p$ plane when $\alpha=0$. The dotted lines
denote $\lambda_p = -2$. Poles on the real axis to the right of this
line correspond to tachyons. There is a branch cut at $\lambda_p =
-17/4$ and the thick line represents the branch cut. There is always a
massless graviton pole at $\lambda_p =-2$. The diagram on the
left is for $\beta > 0$, when there is a single ghost pole. As $\beta$
decreases, this pole moves to the right and another pole emerges from 
the branch cut. This new pole
corresponds to a massive particle and appears in the second diagram,
which is for $\beta = 0$. The final diagram is for $\beta < 0$, when
the ghost pole crosses through $\lambda_p = -2$ and becomes a tachyon.}
\label{fig:poles}
\end{figure}

We shall start by considering the case $\alpha = \beta = 0$, for which
there are two other poles in our propagator, one at $p \approx -1.48$
and the other at $p \approx -0.054$. 
The former has $\lambda_p \approx -17/4$ (but is not quite on the
cut) and has positive residue, the latter has $\lambda_p \approx
-2.16$ and negative residue. The behaviour of $F(p,0,0)$ is plotted in
figure \ref{fig:FR0}. It is easy to show that signs of the residues of $F^{-1}$
with respect to $\lambda_p$ are given by the slope of $F$ as it passes
through $0$. The positions of the poles are shown in figure
\ref{fig:poles}.
Changing the value of $\beta$ (still with $\alpha=0$) changes the
position and nature of these poles. As $\beta$ is made more positive,
the pole with $p \approx -1.48$ gets absorbed into the branch cut and
the ghost moves towards $p = -1$ (i.e. $\lambda_p = -4$). As $\beta$ is
made more negative, the pole with $p \approx -1.48$ moves towards
$p=-1$ while the other pole moves to positive $p$ (i.e., $\lambda_p >
-2$), with its residue changing sign as it crosses $p=0$. 
This pole corresponds to a tachyon. Recall that tachyons were also
present in flat space for sufficiently negative $\beta$. 
In order for tachyons to be absent in flat space, 
we had to choose $\beta > \log 2 - 1$. We have roughly the same
restriction on $\beta$ in order to avoid spin-2 tachyons in de Sitter space. 
We shall therefore exclude the case $\beta < \log 2 - 1$ as unphysical. 

Now consider the effect of turning on $\alpha < 0$. This has no effect
on the pole at $\lambda_p = -2$, so the massless graviton remains. 
If $\beta = 0$, then the two other poles move together as $\alpha$ decreases 
and eventually coalesce into a
double pole. This splits into a pair of complex conjugate poles that
move off into the complex $\lambda_p$ plane. For $\beta > 0$ then there
is generally only one pole present (in addition to the graviton pole)
when $\alpha = 0$. As $\alpha$ is
decreased, an additional pole (with positive residue) emerges from the
branch point and moves towards the ghost pole, eventually coalescing
with it. This then splits into a pair of complex conjugate poles. If
$\beta < 0$, then the two poles again move together, coalesce and then
become a pair of complex conjugate poles. In all cases, the effect
of making $\alpha$ more negative is similar to the effect of increasing $\rho$
in the flat space propagator, i.e., pathologies such as ghosts and
tachyons move off into the complex plane. When $\beta$ is large, the
poles becomes complex for $\alpha < -\beta/8$,
so no fine tuning of the ratio $\alpha/\beta$ is involved.

\subsection{Complex poles}

We have seen how ghost poles can be moved off the real axis, becoming
a pair of complex conjugate poles. The interpretation of such a pair
of poles has been reviewed by Coleman \cite{coleman:69}. The presence
of complex conjugate poles with (complex) masses given by $m = a \pm
ib$ with $b>0$ implies causality violation at lengths or times of the
order of $1/\sqrt{b}$. For Tomboulis' flat space propagator, we have
$b \sim R^{-1}$, so one expects causality to be violated at a length
scale of the order of $R$, which is roughly $N$ times the Planck
length. Unless $N$ is enormous, this is far less than any scale probed
by particle physics experiments so such causality violations are
unobservable\footnote{In fact, these effects might be
smaller than the effects of the gravitational field of subatomic
particles, which would also lead to modifications of causality through
tilting of light cones.}, as noted by Tomboulis.

For our de Sitter propagator, the complex poles again have $b \propto
R^{-1}$. If $|\alpha|$ is large then $b \propto \sqrt{-\alpha} R^{-1}$, 
so causality violation occurs on a time scale
$R/\sqrt{-\alpha}$. If $|\alpha|$ is not large then causality violation
occurs on a time scale $R$. This is much smaller than scales probed 
in experiments, but may have observational
consequences in the CMB since $R$ is the Hubble time, and therefore
the time scale for microphysics during inflation. However, we shall
see in the next section that observations suggest that $|\alpha|$ 
is of order $10^9$, so causality violation occurs on a time scale much 
shorter than the Hubble time and is therefore completely unobservable. 

\sect{Lorentzian two-point correlators}

\label{sec:cont}

In this section we will show how the scalar and tensor propagators
on the four sphere instanton uniquely determine the primordial CMB
perturbation
spectrum in Lorentzian closed de Sitter space. The two-point
correlators
in the Lorentzian region are obtained directly from the Euclidean 
propagators by analytic continuation. 
We refer the reader for the details of this calculation
to our previous paper \cite{hawking:00}, where we described 
the analytic continuation of the graviton correlator in a
Randall-Sundrum version of the Starobinsky model. 
The techniques to perform these calculations were developed in 
\cite{gratton:99,hertog:99}.

\subsection{Scalar propagator}

We have the Euclidean correlator \ref{eqn:scor}
as an infinite sum over real $p$,
where $p$ labels the level of the four sphere scalar harmonics. 
Although this is a convenient
labelling to study their analytic structure, the eigenspace of the 
Laplacian on de Sitter space suggests that the Lorentzian propagator is most
naturally expressed in terms of an integral over real positive
$p' = i(p + 3/2)$, corresponding to scalar harmonics of the Lorentzian
Laplacian with eigenvalue $\lambda_{p'} = ({p'}^2 + 9/4)$.   
We must therefore first analytically continue our result for the
propagators into the complex 
$p$-plane before continuing to Lorentzian signature.
In terms of the label $p'$, the Euclidean scalar correlator
\ref{eqn:scor} becomes
\be
\label{eqn:escor2}
 \langle \psi (\Omega) \psi (\Omega') \rangle = -\frac{32 \pi^2
R^4}{3 | \alpha |N^2}  
\sum_{p'=5i/2}^{+i\infty}
 \frac{W^{(p')}(z(\Omega,\Omega'))}{ p'^2 +9/4 -m^2},
\ee
with
\be
W^{(p')} (z) = \frac{5ip'(p'^2+1/4)}{3\pi^2} \ _2 F_1 (3/2 +ip',3/2 -
ip',2,1-z)
\ee
and $z=\cos^2 (\mu /2)$. 
This biscalar is analytic in the upper half $p'$-plane.
The coefficient of the biscalar
is also analytic in the upper half plane 
apart from a simple pole at $p' = \Lambda_{t}$, where
\be
 \Lambda_t = i \sqrt{\frac{9}{4} - m^2}.
\ee 
This pole corresponds to the tachyon.
Notice that the sum in equation \ref{eqn:escor2} starts at $p'=5i/2$
because we have projected out the negative homogenous mode, which
should be regarded as part of the background \cite{hawking:83}.

Knowing the analytic structure of the correlator,
we are able to write the sum \ref{eqn:escor2} as an integral along
a contour ${\cal C}_1$ encircling the points $p' = 5i/2,7i/2,..ni/2$, 
where $n$ tends to infinity. This yields
\be \label{escor3}
\langle  \psi(\Omega) \psi(\Omega ')\rangle
 = \frac{16i \pi^2 R^4}{3| \alpha | N^2}
\int_{{\cal {C}}_1} dp' \frac{(\tanh p'\pi) 
W^{(p')}(\mu)} { p'^2 + 9/4 -m^2}.
\ee
The contour ${\cal C}_{1}$ can be distorted to run along the real 
$p'$-axis. Apart from the tachyon pole, we encounter two extra
poles at $p'=3i/2$ and $p'=i/2$ in the $\tanh p' \pi$ factor. 
The $p'=3i/2$ pole corresponds to the negative homogenous mode that we
have projected out in the Euclidean correlator. 
On the other hand, $W^{(i/2)}(\mu) =0$ 
so the pole at $p'=i/2$ does not contribute to the propagator.
The contribution from the closing of the contour 
in the upper half $p'$-plane vanishes. Hence our final result for the
Euclidean correlator reads
\be \label{int}
\langle \psi (\Omega) \psi (\Omega') \rangle =  
\frac{16i \pi^2 R^4}{3 | \alpha | N^2}  
\left[ \int_{-\infty}^{+\infty}
dp' \frac{(\tanh p' \pi)  W^{(p')}(z)}{p'^2 + 9/4 -m^2}
 -\frac{\pi i}{\Lambda_t}
(\tanh \Lambda_{t} \pi ) W^{(\Lambda_{t})}(z) 
+ \frac{10i}{m^2 \pi^2} \right].
\ee
Finally one can rewrite \ref{int} as an integral from $0$ to $\infty$,
over the eigenspace of the Lorentzian Laplacian, and the two discrete 
contributions from the tachyon pole and the homogenous mode.
The tachyon contribution grows exponentially for timelike intervals.
However, the relevant propagator for computing the CMB anisotropies 
is the Feynman propagator, which should be bounded both to the past 
and future. Therefore, the propagator that we have obtained by analytic
continuation from the four sphere does
not obey the appropriate boundary conditions.
In order to obtain the two-point
function that describes the correlations in the primordial
density fluctuation spectrum, we change the contour of integration
so as to exclude the contribution from the tachyon pole.
We then obtain the Lorentzian Feynman scalar propagator,
\be
\langle \psi (x) \psi (x') \rangle =  
-\frac{32 \pi^2 R^4}{3| \alpha | N^2}  
\left[ \int_{0}^{+\infty}
dp' \frac{(\tanh p' \pi)  W^{L(p')}(z(x,x'))}{p'^2 + 9/4 -m^2}
+ \frac{10}{m^2 \pi^2} \right].
\ee
The Lorentzian biscalar $W^{L(p')}$ differs 
from $W^{(p')}$ only by a factor of $-i$ 
and $(\tanh p' \pi)  W^{L(p')} (z)$ equals the sum of
the degenerate scalar harmonics on closed de Sitter space with eigenvalue
$\lambda_{p'} = (p'^2 +9/4)$ of the Laplacian.
For spacelike separations, we have $z=\cos^2 (\mu /2)$, where
$\mu (x,x')$ is the geodesic distance between $x$ and $x'$. 
The correlator for timelike 
intervals is obtained by setting $\rho = \pi /2 -it$, where $\rho$ is
the polar angle on the four sphere. For a purely timelike separation,
this gives $z = \cosh^2 ((t-t')/2)$. 

\subsection{Tensor propagator}

The principles of the continuation of the 
tensor propagator \ref{eqn:tensorcorrelator}
are the same, but the calculation is more complicated.
We refer the interested reader to our previous paper
\cite{hawking:00} for the technical details.
The differences between \cite{hawking:00} and the present paper
are that we now have included the effect of the finite
$R^2$ counterterm, we have kept $\beta$ in the coefficient of the
third counterterm arbitrary and we now treat the discrete poles in the
propagator more carefully.

In \cite{hawking:00} it was shown
that the bitensor $W_{iji'j'}^{(p')}(\mu)$
can be unambiguously extended
as an analytic function into the upper half $p'$-plane.
In addition, from subsection \ref{subsec:tensorprop} we know that its
coefficient $F(-ip'-3/2,\alpha,\beta)^{-1}$
is analytic, 
apart from a simple pole at $p'=3i/2$, corresponding to the massless graviton 
in de Sitter space, and a pair of poles with complex masses 
$\Lambda_1$ and $\Lambda_2=-\bar \Lambda_1$ (we are assuming that
$\alpha < - \beta/8$ so that there are complex poles instead of a ghost).
These poles always occur in the upper half $p'$-plane.

Writing the sum in equation \ref{eqn:tensorcorrelator} 
as a contour integral yields 
\be \label{etcor}
\langle  \theta_{ij}(\Omega) \theta_{i'j'}(\Omega ')\rangle
 = -\frac{64 i\pi^2 R^4}{N^2}
\int_{{\cal {C}}_1} dp' \tanh p'\pi
W_{\ iji'j'}^{(p')}(z) G(p',\alpha,\beta)^{-1}
\ee
where
\ba
G(p',\alpha,\beta ) & = & F(-ip'-3/2,\alpha,\beta)\nonumber\\ & = &
p'^4 -4ip'^3 -p'^2/2 -5ip' -3/16 +(p'^2 + 9/4)[4 \alpha + (p'^2+1/4) \times 
\nonumber\\
& & 
(\psi(-ip'/2 +5/4) + \psi(-ip'/2+7/4) -\psi (1) -\psi (2) +2\beta)]. \nonumber
\ea
As we deform the contour towards the real axis we encounter, apart
from the poles mentioned above, two extra poles in the
$\tanh p' \pi$ factor. However, as explained in detail in 
\cite{hawking:00}, they do not contribute to the tensor fluctuation spectrum.
The contribution from the closing of the contour 
in the upper half $p'$-plane vanishes.
Using $ G(-\bar{p}',\alpha,\beta)= \bar{G} (p',\alpha,\beta)$, 
one can again rewrite the remaining integral over the real axis 
as an integral from $0$ to $\infty$. 
The continuation of $z(x,x')$ for timelike intervals
is the same as for the scalar two-point function. 
We then obtain for the Lorentzian tensor propagator,
\ba\label{lcor}
\langle  \theta_{ij}(x) \theta_{i'j'}(x')\rangle
& = & \frac{128 \pi^2 R^4}{N^2} \left\{ \int_{0}^{+\infty} dp'
 (\tanh p'\pi) W_{\ iji'j'}^{L(p')}(z)\, \Re ( G(p',\alpha,\beta)^{-1})
\right. \nonumber\\
& & \left. 
-\pi {\bf R}_{ij i'j'}(z)
- 2 \pi \Re \left[ (\tanh \Lambda_1 \pi)  W_{\
iji'j'}^{(\Lambda_1)}(z)  \, {\bf R}_{(\Lambda_1)}  \right]
\right\}.
\ea
In the integral, $ (\tanh p' \pi)
W_{\ iji'j'}^{L(p')}(z(x,x') )$ can be identified with
the sum of the degenerate rank-two tensor harmonics on closed de Sitter space
with eigenvalue  $\lambda_{p'}= ({p'}^2 +17/4)$ of the Laplacian.
The integrand vanishes as $p' \rightarrow 0$, so the correlator is
well-behaved in the infrared. 

The first term in equation \ref{lcor}
represents the continuous tensor fluctuation spectrum.
The second term describes the massless graviton with ${\bf R}_{ij
i'j'}(z)$ defined as the residue at $p'=3i/2$ of
\be
 W^{(p')}_{\ iji'j'}(z) \,\frac{\tanh p' \pi}{G (p',\alpha, \beta ) }.
\ee
The third term in \ref{lcor} is the combined contribution from the 
complex poles, with 
${\bf R}_{(\Lambda_{1})}$ denoting the residue of
$G(p',\alpha,\beta)^{-1}$ at $p' = \Lambda_1$. For large $| \alpha | $
this mode grows exponentially, implying that the analytically continued
propagator does not obey the boundary conditions for the Feynman propagator.
This can be remedied by changing the contour of integration to exclude
the contribution from the complex poles, giving the correct
propagator for two-point tensor correlations in the
microwave background:
\be
\langle  \theta_{ij}(x) \theta_{i'j'}(x')\rangle
 =  \frac{128 \pi^2 R^4}{N^2} \left[\int_{0}^{+\infty} dp'
 (\tanh p'\pi) W_{\ iji'j'}^{L(p')}(z)\, \Re ( G(p',\alpha,\beta)^{-1})
- \pi {\bf R}_{iji'j'}(z) \right].
\ee
If $|\alpha|$ is large then the tensor propagator is proportional to
$(|\alpha | N^2)^{-1}$. 
At large $p'$ the tensor propagator behaves like $(p'^4 \log p')^{-1}$, just
as the Euclidean correlator \ref{eqn:tensorcorrelator}.
This is in contrast to the usual $p'^{-2}$ behavior of the
graviton propagator for de Sitter space with a cosmological constant.

\sect{Observational constraints}

\label{sec:observe}

\subsection{Duration of inflation}

The Starobinsky instability in four dimensions 
has been analyzed carefully by Vilenkin \cite{vilenkin:85}. 
He showed that the scale factor grows exponentially until
\be
 t = t_* \sim \frac{6 H_0}{M^2} (\gamma-1),
\ee
where, for our model, the parameters $H_0$ and $M$ are given by
\be
 H_0 = R^{-1}, \qquad M = \left(\sqrt{-2\alpha} R \right)^{-1}.
\ee
The parameter $\gamma$ is related to the initial perturbation from the
exact de Sitter solution
\be
 \gamma = \frac{1}{2} \log (2/\delta_0),
\ee
where
\be
 \delta_0 = \frac{H_0 - H}{H_0},
\ee
is the perturbation of the Hubble parameter $H=\dot{a}/a$ at time
$t=0$. If $\delta _0< 0$ then the solution eventually becomes singular
\cite{starobinsky:80}, at least if one neglects spatial curvature
(which should be a good approximation if there is a lot of inflation).
We shall therefore restrict ourselves to $\delta_0 > 0$. 

For $t < t_*$, there is exponential growth with Hubble parameter
$H_0$. The number of e-foldings of inflation during this phase is therefore
\be
 N_1 = \frac{6H_0^2}{M^2} (\gamma-1).
\ee
For our values of $H_0$ and $M$, this gives
\be
 N_1 = -12 \alpha (\gamma-1).
\ee

For $t > t_*$, there is a phase of slow-roll inflation in which the 
Hubble parameter changes from $H_0$ to $M$. The number of e-foldings
of inflation during this phase is \cite{vilenkin:85}
\be
 N_2 = -12 \alpha \log \cosh 1  \approx - 2.26 \alpha.
\ee
The slow-roll phase lasts until $t \sim 6 \gamma H_0/M^2$. Once this
phase ends, the universe enters a matter dominated era in which the
scale factor behaves as \cite{starobinsky:80, vilenkin:85}
\be
 a(t) \propto t^{2/3} \left( 1+ \frac{2}{3 M t} \sin Mt + {\cal
O}(t^{-2}) \right).
\ee
The oscillations in the scale factor can drive particle production and
reheating. 

Vilenkin used the Wheeler-DeWitt equation to obtain an estimate for
$\delta_0$. Using his results, we obtain
\be
 \delta_0 \sim \frac{1}{\sqrt{2}{N}}, \qquad N_1 = -12 \alpha (\log N - 1).
\ee
Quantum cosmology therefore predicts $\gamma \gg 1$. So far, the only
restriction on $N$ is that $N$ must be large enough for our
AdS/CFT calculation to be valid. This implies that $\log N$ is not
close to $1$, so taking $\alpha < -5$ makes $N_1$
sufficiently large to solve the horizon and flatness problems. 

Our correlation functions for metric perturbations were
calculated assuming a four sphere (or de Sitter) background. 
The present day horizon size left the horizon about fifty e-folds
before the end of inflation. Hence the long-wavelength
temperature fluctuations
in the microwave sky carry the imprint of the first expansion phase
provided $N_2 < 50$, which is true if $\alpha > -20$.
Because our correlation functions for metric perturbations were
calculated assuming a four sphere (or de Sitter) background,
the predicted spectrum can then be directly compared with observation.
However, our results will be modified for modes that left the horizon 
during the slow-roll phase, when the background is not exactly de
Sitter. 
Therefore, if $\alpha \leq -20$ then it would be necessary to do a
calculation based on an scalar/vector/tensor decomposition on the
{\it three} sphere in order to enable us to
evolve the spectrum through the instability and predict 
in detail the CMB fluctuation spectrum.

\subsection{Amplitude of perturbations}

In order to compare our results with observations, we should first
render the propagators dimensionless by dividing by $R^4$. 
The correlators are
then functions of $p$ divided by $N^2$. Long wavelength
perturbations are insensitive to what happens after inflation, so
these can be directly compared with observation. For the tensors, long
wavelength perturbations correspond to modes on the four
sphere\footnote{
We should really be studying the Lorentzian correlators here. However,
the overall amplitude of the Lorentzian and Euclidean propagators is
the same.}
with $p=2$. The amplitude of the fluctuations can be obtained from the
correlator:
\be
 \theta_{ij}/R^2 \sim \left(\frac{128 \pi^2}{N^2
 F(2,\alpha,\beta)}\right)^{1/2}.
\ee
In order to agree with observations this should not exceed
$10^{-5}$, which requires
\be
 N^2 (250 + 240 \beta - 40 \alpha) > 10^{13}.
\ee
Since we are assuming $N$ is large, the obvious way to satisfy this
inequality is to take $N = {\cal O} (10^5)$. However, this implies that the
number of fields present is $11 N^2  = {\cal O}(10^{11})$, which seems to
contradict present day observations\footnote{However, it is possible
that these fields may have masses large compared to the scale probed 
in colliders, i.e., $m \gg 1 \, {\rm TeV}$, but
small compared with the scale at which inflation takes
place, $m \ll 10^{-5} m_{pl}$. Such fields would be effectively massless
during inflation but unobservable today.}. Instead, we could take $N^2
\beta$ to be of order $4 \times 10^{10}$ or $N^2 |\alpha|$ to be of
order $2 \times 10^{11}$. The former corresponds to taking
the coefficient of the Weyl squared term in the action to be of order
$10^7$ and the latter corresponds to taking the coefficient of the
$R^2$ counterterm to be of order $10^8$. 

Note that if we take $\beta$ to be large
then we would also have to take $\alpha$ to be large in order to avoid
ghosts in the tensor propagator. Therefore the most natural choice is
probably to take just $\alpha$ to be large. Note that suppression of tensor
perturbations through a Weyl squared counterterm (i.e. taking $\beta$
large) was not mentioned in \cite{starobinsky:80, vilenkin:85} since
this counterterm does not affect the coefficients $a,c,d$ in the trace
anomaly. 

Turning to the scalar perturbations, we see that these can also be
suppressed by taking $N^2 |\alpha|$ to be large. 
Changing $\beta$ does not affect the scalars. 
Our scalar correlator suggests that taking
$N^2 |\alpha|$ to be of order $2 \times 10^{11}$ should bring the scalar 
perturbations within observational bounds. 

We conclude that if $N^2 |\alpha|$ is of order $2 \times 10^{11}$ then 
we can bring metric perturbations within the observational
bounds. $N$ just has to be large enough to justify the large $N$
approximation for the matter fields. For example, we could take $N=10$
and $\alpha = - 2\times 10^9$. 
However, such a large value for $\alpha$ implies that
all modes that we observe today must have left the horizon during the
slow-roll phase of inflation.
Our results for the two-point correlators
will be modified in this case, since we assumed a four sphere
background in our calculation.
However, it is usually the case that the amplitude of 
perturbations is inversely proportional to the horizon radius at which
they left the horizon.
The horizon radius increases during slow-roll
so it seems likely that if $| \alpha |$ is very large 
the amplitude of perturbations will be smaller
than the amplitude obtained above.
This argument is confirmed by the estimates of Vilenkin
\cite{vilenkin:85}. We conclude that taking $N^2 |\alpha| \approx 2 
\times 10^{11}$ will bring the perturbations within observational
bounds, and a far smaller value may in fact be sufficient.

A coefficient of order $10^8$ in the action is large, but this is 
essentially the same fine-tuning problem that also appears in all
scalar field models of inflation. In these scenarios, 
matching the amplitude of perturbations to COBE typically requires 
a fine-tuned parameter in the action of ${\cal O} (10^{-12})$.

Note that taking $|\alpha|$ to be very large implies that causality
violations during inflation occur on a time scale much shorter than
the Hubble time, so they would not have had a significant effect on
microphysics. One might worry that taking $|\alpha|$ to be large would
imply significant deviations from Einstein gravity today, arising from
the higher derivative $R^2$ term in the action. In flat space, the
only effect of this term is to introduce a scalar field with mass
given by equation \ref{eqn:flatscalarmass}. If we take $N=10$ and
$|\alpha|$ of order $10^9$ then this scalar has mass $M \approx 10^{-6}
m_{pl}$, which is far too massive to be observed nowadays.

\sect{Short distance physics}

\label{sec:RS}

\subsection{Introduction}

The observational constraints that we have derived do not depend on
the detailed structure of our propagators and could be obtained directly
from the work of Starobinsky and Vilenkin. In this section we shall
consider a new phenomenon revealed by our propagators, namely the
suppression of short distance metric perturbations by matter
fields. This suppression is evident in Tomboulis' flat space propagator
\ref{eqn:tomboulis}, which falls off as $(k^4 \log k^2)^{-1}$ for large
momentum $k$. It is also present in our tensor
propagator\footnote{
Once again, we shall concentrate on the Euclidean propagators in the
section. The Lorentzian propagators exhibit similar short distance
behaviour.}
, equation \ref{eqn:tensorcorrelator}, which falls off as
$(p^4 \log p)^{-1}$ at large $p$. This behaviour has not been
discussed in previous studies of the Starobinsky model because these
have neglected the non-local part of the matter effective action.

Inflation acts as a ``cosmic magnifying glass'' by blowing up
microscopic physics to macroscopic scales. It is often assumed that this
might lead to some characteristic signature in the CMB of new physics
at short distances, e.g., extra dimensions. Our results appear to
contradict this inflationary dogma, because they show that at small
scales, matter fields will completely drown out the effects of any new
gravitational physics. In this section we shall illustrate this
phenomenon by comparing our results with the results for a model with 
an extra dimension, namely the Randall-Sundrum (RS) \cite{randall:99} 
version of the Starobinsky model.

\subsection{Randall-Sundrum model}

The RS model consists of a five dimensional spacetime with negative 
cosmological constant, and a thin positive tension domain wall whose 
tension is fine tuned to cancel the effect of the bulk cosmological 
constant. The ground state solution of this model is a Poincar\'e
symmetric domain wall separating two regions of AdS. 
In the RS version of the Starobinsky model, we simply add a $U(N)$
Yang-Mills theory to the worldvolume of the domain wall. This model
was extensively discussed in our previous paper \cite{hawking:00}. For
related work, see \cite{nojiri:00a, nojiri:00b, nojiri:00c,
anchordoqui:00}. The (Euclidean) action is
\be
 S=S_{bulk} + S_{brane},
\ee
where
\be
 S_{bulk} = -\frac{1}{16 \pi G_5} \int d^5 x \sqrt{g} \left(R +
\frac{12}{l^2} \right) - \frac{1}{8 \pi G_5} \int d^4 x \sqrt{h} [K]^+_-, 
\ee
\be
 S_{brane} = \frac{3}{4\pi G_5 l} \int d^4 x \sqrt{h} + W[{\bf h}],
\ee
where $g_{\mu\nu}$ denotes the five dimensional bulk metric and
$h_{ij}$ the metric induced on the domain wall, $l$ is the radius of
the AdS solution, $W$ is the generating functional of the Yang-Mills
theory on the domain wall and $[K]^+_-$ is the discontinuity in the trace 
of the extrinsic curvature at the domain wall\footnote{See
\cite{chamblin:99} for an explanation of why this term is required.}. 

There are two simple solutions of the equations of motion for this
model. Since the trace anomaly vanishes in flat space, a Poincar\'e
symmetric solution still exists. However, on a domain wall with de
Sitter geometry, the trace anomaly acts
like an extra contribution to the tension which permits a 
self-consistent de Sitter solution to the equations of motion. 
The Euclidean version of this is a
spherical domain wall separating two balls of AdS. The radius $R$ of
the domain wall is given by \cite{hawking:00}
\be
\label{eqn:radsol}
 \frac{R^3}{l^3} \sqrt{\frac{R^2}{l^2}+1} = \frac{N^2 G_5}{8 \pi l^3} +
\frac{R^4}{l^4}.
\ee
The metric in each bulk region is pure AdS:
\be
 ds^2 = l^2 (dy^2 + \sinh^2 y d\Omega_4^2),
\ee
for $0 \le y < y_0$. The domain wall at $y=y_0$, where $y_0$ is given 
by $R = l \sinh y_0$.

The RS model can be interpreted using the AdS/CFT correspondence as
four dimensional gravity coupled to a Yang-Mills theory with an 
ultraviolet cut-off \cite{gubser:99, hawking:00}. The Yang-Mills theory is
two copies of the ${\cal N}=4$ $U(N_{RS})$ super Yang-Mills theory
with $N_{RS}$ given by
\be
 \frac{l^3}{G_5} = \frac{2N_{RS}^2}{\pi}.
\ee
We shall refer to this dual Yang-Mills theory as the RS CFT in order
to distinguish it from the theory on the domain wall. The Newton
constant in four dimensions is given by the RS value $G_4 = G_5/l$.
The four dimensional dual of the RS model with a $U(N)$ CFT on the
domain wall is four dimensional gravity coupled
to both the RS CFT {\it and} the $U(N)$ CFT. These two CFTs are rather
different in that the former has an ultraviolet cut-off (so its effective
action does {\it not} behave as $p^4 \log p$ at large $p$) whereas the
latter does not. The effective action of the RS CFT is proportional to
$N_{RS}^2$, while the effective action of the other CFT is
proportional to $N^2$. This implies that the effects of the RS CFT
should be negligible when $N \gg N_{RS}$. This is confirmed by
expanding equation \ref{eqn:radsol} in powers of $N/N_{RS}$. At
leading order, one recovers the four dimensional result \ref{eqn:Rdef}. 
Note that $N \gg N_{RS}$ implies $R \gg l$, i.e., the domain wall is
large compared with the AdS length scale.

\subsection{Brane-world perturbations}

The RS model is a short distance modification of gravity. For length
scales much greater than the AdS length $l$, four dimensional gravity
is recovered. However, at shorter distances gravity becomes five
dimensional. One might expect this to lead to a characteristic signal
in the CMB. This turns out not to be true when the Yang-Mills theory
is included on the domain wall. The reason is simple: at short
distances, the matter contribution to the graviton propagator
completely dominates the contribution from the four or five
dimensional Einstein-Hilbert action. One might think that this effect
is peculiar to our model of anomaly driven inflation, and would not
occur in other models of inflation. However, any model has to take
account of the Standard Model, which contains a large number of
fields. These matter fields will suppress small scale metric
perturbations in the same way as our Yang-Mills theory.

We shall illustrate this effect explicitly by calculating the scalar
and tensor graviton propagators for anomaly driven inflation 
in the RS model. Our method will be the same as above, i.e., we shall
calculate the propagators in Euclidean signature and analytically
continue to Lorentzian signature. The initial quantum state of
perturbations is defined by the boundary condition of regularity on
the Euclidean solution.  In the RS case, this condition of regularity
extends into the bulk. 

This work is an extension of our previous paper \cite{hawking:00},
which contained the first rigorous derivation of cosmological
perturbations in RS cosmology. However, in that paper we only
discussed tensor perturbations and did not include the finite $R^2$
counterterm. Here, we shall include this counterterm and also consider
scalar perturbations. Our method involves integrating out metric
perturbations in the fifth dimension. For alternative approaches to
brane-world cosmological perturbations, see \cite{mukohyama:00,
kodama:00, maartens:00, langlois:00, vandeBruck:00, koyama:00}.

The metric perturbation on the domain wall can be decomposed as in
subsection \ref{subsec:pert}, giving a scalar $\psi(x)$ and a tensor
$\theta_{ij}(x)$. Correlation functions of these quantities can be
calculated by integrating out the bulk metric perturbation, as
explained in \cite{hawking:00}. This is done by splitting the bulk
metric perturbation $\delta {\bf g}$ into a classical part $\delta
{\bf g}_0$ and a quantum part $\delta {\bf g}'$. The classical
part is the solution of the linearized Einstein equation in the bulk
that is regular throughout the bulk and  
matches onto the metric perturbation at the domain wall. The
quantum part vanishes at the domain wall. Performing the path integral
over $\delta {\bf g}'$ gives some determinant $Z_0$ that we shall not
worry about. The classical part simply contributes the bulk
action evaluated on shell:
\be
 \int d[\delta {\bf g}] e^{-S_{bulk}[\delta {\bf g}]} = Z_0
 e^{-S_{bulk}[\delta {\bf g}_0]}.
\ee 
We conclude that the effective action governing metric perturbations
on the domain wall is
\be
 S_{eff} = 2S_{bulk}[\delta {\bf g}_0] + S_{brane}.
\ee
The factor of $2$ is necessary if we regard $S_{bulk}$ as the action
of just one of the bulk regions. $S_{brane}$ is straightforward
to compute using our result for $W$, equation \ref{eqn:CFTgenfun}. The
bulk metric perturbation $\delta {\bf g}_0$ can be obtained from the
results of section \ref{sec:pert} by replacing $\bar{l}$ and $\bar{G}$
by $l$ and $G_5$. It follows that the bulk metric perturbation is
transverse traceless, and the scalar $\psi$ arises from a perturbation
in the position of the domain wall in Gaussian normal coordinates.
$S_{bulk}$ can be obtained from equations 
\ref{eqn:Sgrav0}, \ref{eqn:Sgrav1} and \ref{eqn:Sgrav2} 
since the bulk action in the RS model is exactly
the same as the bulk action for the AdS/CFT correspondence. 

From $S_{eff}$ one can read off the metric propagators.
The Euclidean scalar correlator can be written in position space as
\be
\label{eqn:RSscalarcorrelator}
 \langle \psi (x) \psi (x') \rangle = \frac{32 \pi^2 R^4}{3N^2
 (-\alpha) (4+m^2)} \left[ \frac{1}{-\hat{\nabla}^2 + m^2} -
 \frac{1}{-\hat{\nabla}^2 -4} \right],
\ee
where
\be
 m^2 = \frac{1}{2\alpha} \left(\frac{1+2
 e^{-2y_0}}{1+e^{-2y_0}}\right).
\ee
The tensor correlator is
\be
\label{eqn:RStensorcorrelator}
 \langle \theta_{ij}(x) \theta_{i'j'}(x') \rangle = \frac{128 \pi^2
 R^4}{N^2} \sum_{p=2}^{\infty}
 W^{(p)}_{ij i'j'}(x,x') F(p,y_0, \rho, \alpha)^{-1},
\ee
where
\ba
\label{eqn:RSFdef}
 F(p,y_0,\alpha,\beta) &=& e^{y_0} \sinh y_0
 \left(\frac{f_p'(y_0)}{f_p(y_0)} + 4\coth y_0 -6 \right) + \Psi(p) 
 \nonumber \\
 &+& 2 \beta p(p+1)(p+2)(p+3) - 4\alpha p(p+3). 
\ea
Recall that $y_0$ is defined by $R=l \sinh y_0$. We have used equation
\ref{eqn:radsol} to write $l^3/G_5$ in terms of $R$. $f_p$ is defined in
equation \ref{eqn:fdef}. Equation \ref{eqn:RSFdef} was derived in
\cite{hawking:00} but the term involving $\alpha$ was not included.
In comparing our propagators in the RS model with those of the four
dimensional model, we first render them dimensionless by dividing 
by $R^4$.

The scalar correlator for the RS model is very similar to that of the
four dimensional model, as given by equation
\ref{eqn:scalarcorrelator}. The only difference is the
$y_0$-dependence of the tachyon mass $m^2$. 
As $y_0 \rightarrow \infty$, the four dimensional
value is recovered. This is to be expected since, in this limit, $R/l
\rightarrow \infty$, which implies $N/N_{RS} \rightarrow \infty$ using
equation \ref{eqn:radsol}. We have already discussed how the RS
corrections are expected to be negligible when $N \gg N_{RS}$. Note
that as $y_0$ increases from $0$ to $\infty$, $m^2$ just changes
monotonically by a factor of $2/3$.  

The analytic structure of the RS tensor propagator is very similar to
the four dimensional case. There is always a pole at $p=0$: this is
the massless graviton of the RS model\footnote{This pole was mistakenly
identified as gauge in \cite{hawking:00}.}. Other poles behave as
discussed in subsection \ref{subsec:tensorprop}. 

The tensor propagator appears to exhibit more interesting dependence
on $y_0$. The first term in equation \ref{eqn:RSFdef} arises from the
gravitational part of the action, so this is where differences between
a RS model and the four dimensional model show up. As $y_0 \rightarrow
\infty$, the first term tends to $p^2 + 3p + 6$, in agreement with the
four dimensional result (equation \ref{eqn:Fdef}). For very small
$y_0$, the first term is $p + 6$. If $y_0$ is held fixed but large
then the first term grows quadratically with $p$ as $p$ is increased 
but eventually becomes linear for sufficiently large $p$, 
corresponding to gravity becoming five dimensional at short distances. 
Thus the difference between a RS model and four dimensional gravity
might be expected to show up in $1/p$ behaviour in the tensor
propagator at large $p$, rather than the usual $1/p^2$
behaviour. However, this neglects the effects of the matter fields,
which are given by the other terms in equation \ref{eqn:RSFdef}. At
large $p$, $\Psi(p)$ grows like $p^4 \log p$ and completely dominates
the first term. Therefore, at large $p$ the tensor propagator behaves 
like $(p^4 \log p)^{-1}$ irrespective of whether 
one is considering a RS model or
four dimensional gravity. The differences between the RS model and four
dimensional gravity are drowned out by the damping effect of matter
fields at short distances, rendering them unobservable.

\begin{figure}
\begin{picture}(0,0)
\put(25,0){\tiny{$-1.5$}}
\put(70,0){\tiny{$-1.0$}}
\put(120,0){\tiny{$-0.5$}}
\put(177,0){\tiny{$0.0$}}
\put(227,0){\tiny{$0.5$}}
\put(277,0){\tiny{$1.0$}}
\put(15,10){\tiny{$-5$}}
\put(20,38){\tiny{$0$}}
\put(20,70){\tiny{$5$}}
\put(18,103){\tiny{$10$}}
\put(18,133){\tiny{$15$}}
\put(18,165){\tiny{$20$}}
\put(18,195){\tiny{$25$}}
\end{picture}
\centering{\psfig{file=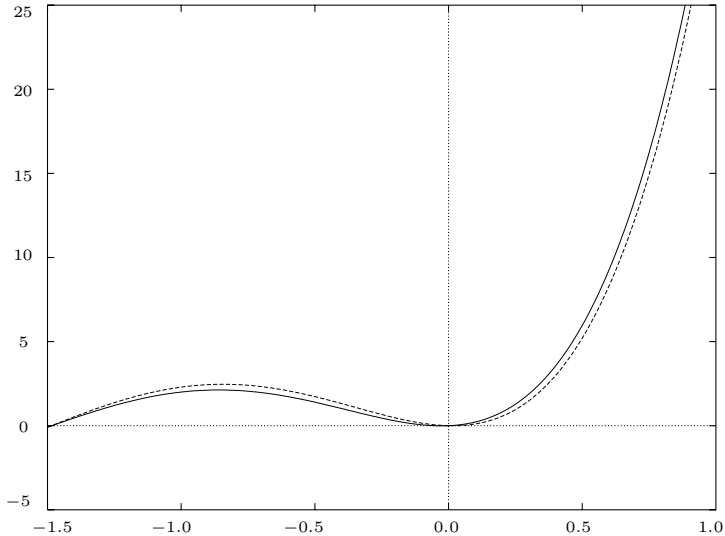,width=4.in}}
\caption{$F(p,y_0,0,0)$ plotted against $p$. The
lower curve on the left (upper curve on the right) is for $R \gg l$,
when four dimensional gravity is recovered. The other curve is for
$R=l$, when the RS corrections might be expected to be large. 
However, they clearly have very little effect.}
\label{fig:compare}
\end{figure}

RS corrections are expected to be important at distances of order
$l$. If we take $R=l$ then all the tensor harmonics have wavelengths
smaller than $l$, not just the large $p$ ones. Therefore, one might
expect RS corrections to be important at small $p$ for such a small
domain wall. Surprisingly, this turns out not to be the case, as shown
in figure \ref{fig:compare}.
This surprising behaviour can be understood in the four dimensional
dual picture. Taking $R=l$ corresponds to $N^2 \approx 6.4 N_{RS}^2$,
so the matter on the domain wall still dominates the effect of the RS
corrections. The RS corrections would be expected to be about as
important as the matter on the wall when $N_{RS} \approx N$, which
corresponds to $R \approx 0.46 l$. In other words, the RS corrections
only become large when the {\it entire domain wall} is smaller than the AdS
radius. 

One might worry that introducing a cut-off into the matter theory
would spoil the damping at large $p$. 
However, if we did have a momentum cut-off
$\Lambda$ then we would need $\Lambda R \gg 1$ in order for field
theory to be valid during inflation, as is always assumed. It
therefore seems appropriate to take $\Lambda \sim m_{pl}$, which
corresponds to $p_{max} \sim N \gg 1$. Figure \ref{fig:compare} shows
that the matter fields dominate the propagator even for quite small
$p$, so introducing a cut-off would have little effect.

\sect{Conclusions}

There is now good observational evidence suggesting that the early
universe underwent a period of inflationary expansion. Most
theoretical models of inflation involve a scalar field rolling down
its potential. The simplicity of such models is attractive but they
have several serious problems. All these models require contrived
initial conditions -- no explanation is given of why the scalar field
was initially displaced from the minimum of its potential\footnote{
Quantum cosmology can answer this question, but only for very contrived
false-vacuum potentials \cite{coleman:80, hawking:82}.}. Secondly, in
order to obtain sufficient inflation and small CMB fluctuations, the
CMB potential has to be highly fine-tuned. Finally, models of scalar
field driven inflation usually disregard the effect of the large
number of other fields in the universe. It is usually argued that the
effect of such fields rapidly becomes negligible during
inflation. However, as we have seen, this is not necessarily true because the
trace anomaly of matter fields provides an additional contribution to
the cosmological constant constant during inflation. 

In this paper, we have argued in favour of Starobinsky's model of
trace anomaly driven inflation \cite{starobinsky:80}
as an alternative to scalar field
driven inflation. In Starobinsky's model, the trace anomaly supports a
de Sitter phase of expansion which is unstable, but can be long
lived. This model is better motivated from the point of view of 
initial conditions since quantum cosmology predicts that the de Sitter 
universe can nucleate semi-classically via a four sphere instanton
\cite{vilenkin:85}. We have seen that this model admits a second
instanton. This can probably be interpreted in a similar way to the
Coleman-de Luccia \cite{coleman:80} instanton, i.e., as describing the
semi-classical decay of the de Sitter phase via nucleation of a pair
of bubbles, each containing an open inflationary universe. Owing to
the lack of an analytic solution for this instanton, 
we have concentrated on the four sphere instanton in this paper.

During the de Sitter phase, particle masses would have been small
compared with the spacetime curvature so matter fields would have been
classically conformally invariant. Moreover, we observe a large number
of fields today and supersymmetry predicts that there should be many
more, so the large $N$ approximation is justified in studies of trace
anomaly driven inflation. This leads to a very attractive way of
calculating the effective action of matter fields during the de Sitter
phase, viz the AdS/CFT correspondence. Using AdS/CFT, we have
presented the first calculation of scalar and tensor metric
propagators for trace anomaly driven inflation, taking full account of
the back-reaction of matter fields.

In order for the de Sitter phase to be unstable, it is necessary for
the coefficient $d = \alpha N^2/(16\pi^2)$ of the $\nabla^2 R$ term in
the trace anomaly to be negative (in our conventions). We therefore
included a $R^2$ counterterm in the action to control this
coefficient. We also took account of the other curvature squared
counterterms. We demonstrated that the amplitude of long wavelength
metric perturbations could be brought within observational bounds at
the expense of fine-tuning of $N^2 |\alpha|$. This fine-tuning is no
worse than required in scalar field driven inflation, 
and agrees with the results of Vilenkin \cite{vilenkin:85}. In fact,
the amount of tuning required may be much less than for scalar field
driven inflation. A more detailed treatment of the slow-roll phase 
would be required to verify this.

One might worry that introducing a $R^2$ counterterm into the action
would lead to observational consequences for, say, solar system
physics. However, the effect of this term in flat space is just to
introduce a scalar field whose mass is of order 
$m_{pl}/(N\sqrt{-\alpha})$. Even though $|\alpha|$ is very large, this
mass is still much too large to lead to observable effects today. For
example, taking $N=10$ and $\alpha$ of order $10^9$ gives a mass of
order $10^{-6} m_{pl}$. 

Our tensor propagator exhibits interesting analytic structure. We have
shown that ghosts can be removed without fine-tuning, although this
introduces a pair of complex conjugate poles. Such poles were studied
long ago and found to correspond to violations of causality. We have
seem that this causality violation occurs on a time scale
$R/\sqrt{-\alpha}$, where $R$ is the Hubble time. This time scale is
much smaller than $R$ when $|\alpha|$ is large enough to bring the
amplitude of metric perturbations within the observational
bound\footnote{
Even if the time scale for causality violation were the Hubble time,
it is not clear that this would contradict cosmological observations
and such violations would certainly not be observable in the laboratory.}. 

At large $p$, the tensor propagator exhibits the behaviour first
discovered for flat space by Tomboulis \cite{tomboulis:77}, namely
suppression of metric perturbations by matter fields. This suppression
does not involve fine-tuning, as required for suppression of
long-wavelength perturbations. The matter fields make the tensor
propagator decay like $(p^4 \log p)^{-1}$ at large wavenumber
$p$. This behaviour would be expected whenever the large $N$ expansion
is valid. Since we observe a large number of matter fields, we have
argued that this suppression should occur even if inflation were not
driven by a trace anomaly. This implies that matter fields damp out
the effects of any short distance modifications of gravity (such as
extra dimensions), rendering them unobservable. We illustrated this
effect by comparing the propagators for trace anomaly driven inflation
in four dimensions and in a Randall-Sundrum model. At large $p$, the
tensor propagators are indistinguishable and at small $p$ they only
differ when the entire domain wall is smaller than the radius of
curvature of the fifth dimension. 

There are many directions in which our work could be extended. 
For example, our use of AdS/CFT has restricted us to a strongly
coupled theory. However, we have argued that our 2-point functions are
independent of the Yang-Mills coupling. Dependence on the coupling
would be expected to show up in higher order correlation functions of
metric perturbations. This implies that these higher order correlation
functions would not be determined by the 2-point functions, so the 
spectrum of CMB fluctuations would exhibit non-Gaussianity.

In the Einstein static universe, the strongly coupled 
Yang-Mills theory exhibits a
confinement/ de-confinement transition at a certain temperature,
corresponding to two different bulk solutions in the
AdS/CFT correspondence \cite{witten:98}. One might therefore wonder
whether there is a bulk solution different from pure AdS which could
have a spherical boundary with an $O(4)$ symmetric metric. If so, then
one might have a phase transition in a cosmological background. This
does not appear possible. To see this, assume that the 
$O(4)$ isometry group of the boundary implies a corresponding
$O(4)$ isometry group in the bulk (we are thinking of a cut-off CFT,
corresponding to a finite boundary). Birkhoff's theorem then implies
that the bulk is (Euclidean) Schwarzschild-AdS. 
However, in order for the instanton to have
spherical topology, the orbits of the $O(4)$ group have to degenerate at two
points on the instanton (the poles) and this is not possible if the
bulk is Schwarzschild-AdS except when the mass parameter vanishes. In
other words, there is a unique solution (pure AdS) in the bulk and therefore no
phase transition. This bulk solution corresponds to a deconfined phase
of the Yang-Mills theory (this is evident from the overall $N^2$
factor in the Yang-Mills effective action).

When one has a choice between several cosmological instantons, one
usually argues that the instanton with the least Euclidean action is
preferred, on the basis that this instanton would give the dominant
contribution to a gravitational path integral. Instantons which are
saddle points, rather than local minima of the action, would not be
viewed as satisfactory. These instantons would possess negative modes,
corresponding to directions in field space along which the action
decreases. Such instantons have been extensively
discussed in \cite{gratton:00}, where it was argued that they may
be interpreted as describing quantum tunneling in an existing
universe, rather than creation of a universe from nothing. Since we
have found two instantons, it would be interesting to examine whether
they have negative modes. This could give support to the idea that the
double bubble instanton describes an instability of the de Sitter
vacuum. 

Any discussion of negative modes presupposes 
the existence of a gravitational path integral. 
This is not well-defined even for
Einstein gravity since it is well-known that the Euclidean
gravitational action is unbounded below. In our case, the presence of
the $R^2$ counterterm with a negative coefficient appears to make matters even
worse. However, it is known that
Einstein gravity coupled to a $R^2$ term can be rewritten as Einstein
gravity coupled to a scalar field \cite{whitt:84} so the situation is 
probably no worse than usual.

We have emphasized that there are two instantons in the Starobinsky
model. However, there is also a third, namely flat space viewed as  
the infinite radius limit of the four sphere. This has
infinitely negative Euclidean action. It might therefore be necessary
to invoke the anthropic principle to explain why an inflationary universe
is nucleated rather than an empty flat universe. The situation is
analagous to false vacuum decay \cite{coleman:80, hawking:82}, 
for which the instanton describing nucleation of a universe in the 
true vacuum state has lower action than the instanton describing 
nucleation of a universe in the false vacuum state. Clearly there is
plenty of scope for future work on understanding the quantum cosmology
of the Starobinsky model.

Our approach was based on decomposing
the metric perturbation into scalar, vector and tensor representations
of $O(5)$, or $O(4,1)$. This made the AdS/CFT calculation relatively
straightforward, but means that our results are only directly
applicable to the initial de Sitter phase, although we have argued that 
the amplitude of metric perturbations should not increase during the
slow roll phase. In order to produce a detailed fluctuation spectrum
that could be compared with observation, it would be necessary to do a
calculation based on a decomposition into scalar, vector and tensor
represenations of $O(4)$ (assuming a closed universe). If the AdS/CFT
calculation could be extended to perturbations around a Euclidean 
background with a general $O(4)$ invariant metric then, by analytic
continuation, one could calculate how the metric propagators evolve
during the slow-roll phase. The perturbations spectrum at the end of
inflation could then be used to predict the detailed spectrum of
temperature fluctuations in the CMB. An $O(4)$ approach would also be
necessary to investigate the double bubble instanton. 

\bigskip

\centerline{{\bf Acknowledgments}}

We are very grateful to Steven Gratton, Hugh Osborn and Neil Turok for useful
discussions. TH is Aspirant FWO, Belgium.

\end{document}